\documentclass[fleqn,10pt]{wlscirep}
\usepackage[utf8]{inputenc}
\usepackage[T1]{fontenc}
\usepackage{authblk}
\usepackage{xspace}  
\usepackage{xcolor}  
\usepackage{siunitx} 
\usepackage{amsmath}
\usepackage{booktabs}
\usepackage{multirow}
\usepackage{subcaption}
\usepackage{placeins}
\usepackage{wrapfig}

\newcommand{\cotwo}{\ensuremath{\ensuremath{\mathrm{CO}_2}}\xspace}
\newcommand{\tcotwoeq}{\ensuremath{\ensuremath{\mathrm{tCO}_2}\mathrm{e}}\xspace}
\newcommand{\gcotwoeq}{\ensuremath{\ensuremath{\mathrm{gCO}_2}\mathrm{e}}\xspace}

\title{Know your footprint -- Evaluation of the professional carbon footprint for individual researchers in high energy physics and related fields}

\author[1,*]{Valerie S. Lang}
\author[1]{Naman Kumar Bhalla}
\author[1,2]{Simran Sunil Gurdasani}
\author[2]{Pardis Niknejadi}

\affil[1]{Albert-Ludwigs-Universität Freiburg, Freiburg, Germany}
\affil[2]{Deutsches Elektronen-Synchrotron DESY, Hamburg and Zeuthen, Germany}

\affil[*]{valerie.lang@physik.uni-freiburg.de}


\begin{abstract}
As the climate crisis intensifies, understanding the environmental impact of professional activities is paramount, especially in sectors with historically significant resource utilisation. This includes High Energy Physics (HEP) and related fields, which investigate the fundamental laws of our universe. As members of the \emph{young High Energy Physicists} (yHEP) association, we investigate the \cotwo-equivalent emissions generated by HEP-related research on a personalised per-researcher level, for four distinct categories: \emph{Experiments}, tied to collaborations with substantial infrastructure; \emph{Institutional}, representing the resource consumption of research institutes and universities; \emph{Computing}, focussing on simulations and data analysis; and \emph{Travel}, covering professional trips to conferences, etc. The findings are integrated into a tool for self-evaluation, the \emph{Know-your-footprint} (Kyf) calculator, allowing the assessment of the personal and professional footprint and optionally sharing the data with the yHEP association. The study aims to heighten awareness, foster sustainability, and inspire the community to adopt more environmentally responsible research practices urgently.
\end{abstract}
\begin{document}

\flushbottom
\maketitle
\thispagestyle{empty}

\section{Introduction}
\label{introduction}

The impact of the atmospheric presence of \cotwo on the ground temperature of the Earth was first predicted in 1896~\cite{Arrhenius}, indicating temperature variations as a function of the amount of \cotwo (at the time denoted as carbonic acid), the time of the year and the latitude. For Europe, the prediction for doubling the atmospheric \cotwo content amounted to an average temperature increase of around \SI{6}{^\circ C}.
In 2023, the measured \cotwo content in the atmosphere has reached nearly \SI{425}{ppm}~\cite{KeelingCurve}, compared to a maximum of about \SI{300}{ppm} over the last \SI{800000} years before industrial age. The \cotwo content in the atmosphere has thus increased by a factor of about 1.4 compared to the maximum, and around 1.9 compared to the mean over the last \SI{800000} years \cite{co2inlast800kyears}.
The average temperature in Europe in 2023 has climbed up by \SI{1.90}{^\circ C} compared to pre-industrial levels (reference years: 1850-1900)~\cite{CopernicusTemperatureData}, which is also used as reference for temperature differences in the following. The year 2023 is in direct competition with the year 2020 which was the warmest year on record in Europe with a \SI{2.07}{^\circ C} increase. Globally, 2023 surpassed the previously warmest year 2016 (\SI{1.32}{^\circ C} increase), reaching an increase of \SI{1.48}{^\circ C}.
Scientific research has thus been drawing a consistent picture for more than 100 years, with continuously refined measurements and scenario predictions. Decisive actions counteracting the changes to the atmosphere and the planetary ecosystem, however, have been and are still lacking.

According to the Sixth Assessment Report (AR6) by the Intergovernmental Panel on Climate Change (IPCC), Summary for Policymakers (SPM)~\cite{IPPCAR6SPM}, \emph{``[i]t is unequivocal that human influence has warmed the atmosphere, ocean and land''}, leading to changes across the climate system that are \emph{``unprecedented over many centuries to many thousands of years''}.
Relevant for changes of the climate are the cumulative \cotwo emissions. Part of the \cotwo emissions are re-absorbed by carbon sinks on land and ocean, but the higher the cumulative \cotwo emissions are, the lower is the fraction of absorption by land and ocean carbon sinks. This results in higher contributions for both the fraction and the absolute amount that remain in the atmosphere. 

The IPCC AR6 SPM estimates historical cumulative anthropogenic \cotwo emissions from 1850 to 2019 to \SI{2390\pm240}{}~Gigatonnes of \cotwo (Gt\cotwo). In order to achieve a likelihood of \SI{67}{\%} to limit human-induced global warming to a maximum increase of \SI{1.5}{^\circ C} (\SI{2.0}{^\circ C}), which are the two thresholds explicitly mentioned in the Paris Agreement 2015~\cite{ParisAgreement} to reduce risks and impacts of climate change, only \SI{400}{Gt\cotwo} (\SI{1150}{Gt\cotwo}) remain to be emitted. This remaining carbon budget is defined, starting from the beginning of 2020 until global net zero \cotwo emissions are reached.

Using the most naive assumption of equal yearly emissions until 2050, i.e. for 30 years, and a static population of 8 billion people, which was reached on 15 November 2022~\cite{UNpopulationData}, the remaining carbon budget per person per year accumulates to \SI{1.7}{t\cotwo} (\SI{4.8}{t\cotwo}) for a maximum temperature increase of \SI{1.5}{^\circ C} (\SI{2.0}{^\circ C}).
In Germany, the average \cotwo emissions accumulate to around \SI{10}{}--\SI{11}{t\cotwo} per person per year~\cite{UBAemissionsGermany, UBAcalculatorRef}, i.e. more than a factor of two above what is required for the scenario limiting global warming to an increase of \SI{2.0}{^\circ C}. This, however, only summarises emissions from private life. For researchers, emissions resulting from research activities need to be added on top. 

One area with historically significant resource utilisation, given the large infrastructures needed for this type of research, is high energy physics (HEP), aiming to understand the fundamental principles of matter and its interactions in the universe. In recent years, efforts to quantify carbon emissions from HEP infrastructures have increased: from the institutional side with the environmental reports by the European centre for particle physics research CERN~\cite{CERN1718, CERN1920, CERN2122} and the German national laboratory DESY~\cite{DESY1921}, first life-cycle assessments of potential future projects~\cite{LCA4ILCandCLIC, FCCfootprint, CEPCfootprint, C3footprint}, and from the individual researchers' side through statements, white papers and reviews~\cite{ALLEA, HECAPplus, SnowmassSustInput, SustFutureAcc, AstroInLowCarbonFuture}. In France, the \emph{Labos 1point5}~\cite{Labos1point5} initiative has prepared a tool to compare the carbon footprints of (mainly French) research institutions through a common framework~\cite{GES1point5}; in the area of astronomy, the carbon footprint of a researcher at the Heidelberg Max Planck Institute for Astronomy was estimated~\cite{AstrMaxPlanckInstEmissions} and the development options for future astronomical research infrastructures investigated~\cite{AstrFutureCarbonEmissions}. While the estimates vary both in numbers and scopes, the tendency is clear: Large-infrastructure research has a significant carbon footprint, which needs to be reduced urgently. 

Reducing emissions requires awareness of the issue which is best achieved if the numbers of carbon emissions can be directly related to one's own life. While some of the above studies already break down the institutional footprint per person, a personalised approach with a carbon calculator is missing.
This study, as part of the \emph{Know your footprint} (Kyf) campaign by the \emph{young High Energy Physicists} (yHEP) association~\cite{yHEPwebpage}, aims to quantify emissions and raise awareness for the professional footprint of researchers in the area of HEP and related fields, and provides a personalised carbon calculator for this purpose. This allows to identify areas requiring most urgent action and indicates possible starting points for every individual researcher to reduce the footprint. While scientific research and development activities contribute only a fraction of the carbon footprint compared to industrial activities, it is crucial for the scientific community to understand and mitigate the environmental impact of their research.

\section{Results}

\subsection{Study setup}
\label{sec:setup}

This paper discusses the data sources, calculations, and results of evaluating the \cotwo-equivalent emissions for researchers in HEP and related areas for yHEP's Kyf campaign and Kyf calculator. The study focusses on German and European research infrastructures as the core of yHEP's activities, but the methodology is applicable worldwide and can be expanded in the future.
In this study, the carbon footprint of researchers is investigated by differentiating between the footprint of private and professional activities. For emissions of private life in Germany, we refer to the \emph{Carbon Calculator} by the German Federal Environment Agency (\emph{Umweltbundesamt} - UBA)~\cite{UBAemissionsGermany, UBAcalculatorRef, UBAcalculator}, with due permission from \emph{KlimAktiv}~\cite{KlimAktiv}, UBA's partner for the Carbon Calculator.

For the professional footprint, we target four distinct categories: \emph{Experiments}, \emph{Institutions}, \emph{Computing} and \emph{Travel}. 
The \emph{experimental} footprint comprises the emissions from (large) research infrastructures, collaborations or research projects, i.e. especially in experimental physics, the experimental setup. The \emph{institutional} footprint summarises the emissions from the home university or research centre of the individual researcher, which should be understood as the institution where the researcher spends the larger part of her/his working time. \emph{Computing} focusses on the individual researcher's footprint of running simulations and data analysis, optionally considering data storage impacts as well. \emph{Travel} covers the individual's professional trips such as to conferences, meetings, or workshops.

The footprints in each category are broken down to the level of the individual researcher so that they can be placed in relation to each other as well as to the individual's emissions for private life. Several options in each category allow adjusting to the researcher's individual situation, although some simplifications are needed to allow for sufficient user-friendliness and manageable scope of the first version of the Kyf calculator by the yHEP association. In some cases, this includes assumptions that cannot be easily validated with the available data. The assumptions made, however, are outlined in the following sections, together with the associated reasoning. Refinements are considered possible in the future. The Kyf calculator is published on the webpage of the yHEP association~\cite{yHEPwebpage}.

The Kyf calculator provides feedback to individual researchers regarding their own professional carbon footprint, and optionally allows to share the data anonymously with the yHEP association. This data will be used by the yHEP association to obtain and later publish an overview of the professional footprint of researchers in HEP and related fields in, and associated to, Germany.
The focus on Germany implies that numbers for the German electricity grid, German institutions, etc. will be used in the calculations, where applicable. Other institutions or interest groups are encouraged to transfer the considerations by yHEP's Kyf campaign to different countries, citing this publication as basis for their considerations. In case of questions, contact can be established with us as authors directly, or with the yHEP association through the webpage~\cite{yHEPwebpage}.

\subsection{Professional footprint example}
\label{example}

\begin{figure}[hbt]
    \centering
    \includegraphics[width=0.59\textwidth]{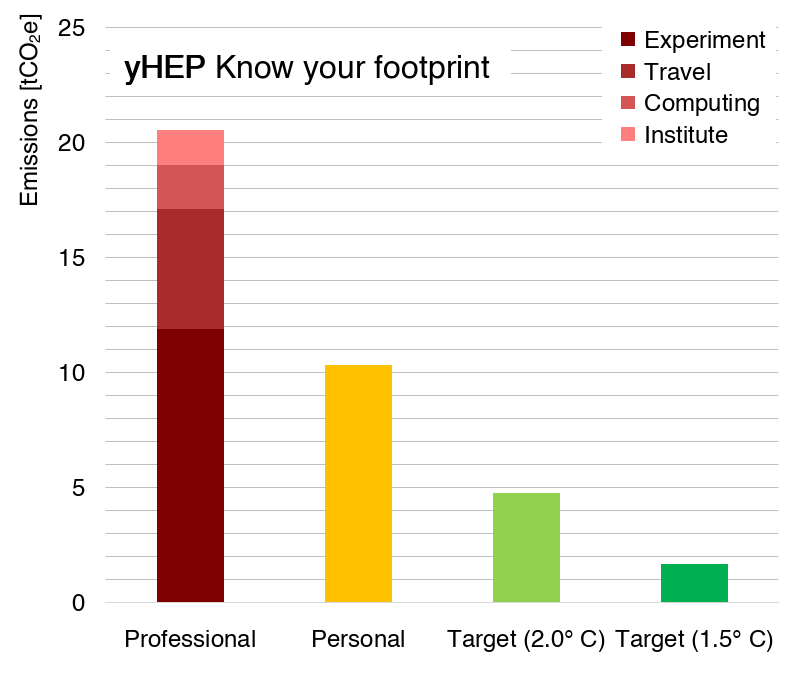}
    \caption{Professional \cotwo footprint of a benchmark doctoral researcher, working on one of the large LHC experiments, employed by a German university, with medium computing usage level and travel as discussed in the text, in comparison with the average personal footprint in Germany, as well as the remaining carbon budget per person per year of \SI{1.7}{\tcotwoeq} (\SI{4.8}{\tcotwoeq}) for a maximum temperature increase of \SI{1.5}{^\circ C} (\SI{2.0}{^\circ C}).}
    \label{fig:Totalfootprint}
\end{figure}

An example professional footprint of an early-career researcher in Germany is calculated for a benchmark doctoral student working on one of the large LHC experiments at CERN, and being employed by a university in Germany, which is supplied with conventional electricity. A medium computing level, where the computing centre also uses conventional electricity, is assumed and travel of two 1-week trips in Germany by train -- for example, the spring meeting by the German Physical Society (DPG) and a national collaboration meeting, one flight travel within Europe for a week -- for example, a conference in Thessaloniki, Greece, and one two-week flight travel across continents -- for example, a summer school in Seoul, South Korea -- during one particular year. The calculations that form the basis of the benchmark researcher's footprint are discussed in detail in Sections~\ref{sec:exp} to \ref{sec:travel}. The combined professional footprint is listed in Table~\ref{tab:ProfFootprint}, and shown in Figure~\ref{fig:Totalfootprint}, in comparison with the personal footprint and the remaining carbon budget per person per year, given the climate warming limitation goals discussed in Section~\ref{introduction}.

\begin{table}[hbt]
\centering
\begin{tabular}{l r}
\toprule
\textbf{Category}   & \multicolumn{1}{c}{\textbf{Emissions [\tcotwoeq]}}    \\
\midrule
Large LHC experiment  & \SI{11.91}{}   \\
University (German electricity mix) & \SI{1.54}{}    \\
Computing (medium)      & \SI{1.91}{}    \\
Travel (2xT, 1xF(E), 1xF(C))   & \SI{5.20}{}    \\
\midrule
Total           & \SI{20.56}{} \\
\bottomrule
\end{tabular}
\caption{Professional \cotwo footprint of a benchmark doctoral researcher. The researcher is assumed to be working on one of the large LHC experiments, employed by a German university, with medium computing usage level and two intra-Germany travels by train (T), one intra-Europe flight travel (F(E)) and one cross-continental flight travel (F(C)) per year.}
\label{tab:ProfFootprint}
\end{table}

The annual professional footprint of the benchmark doctoral researcher amounts to \SI{20.56}{\tcotwoeq}, which can be compared to the average private footprint of \SI{10}{}--\SI{11}{\tcotwoeq} per person per year in Germany~\cite{UBAemissionsGermany, UBAcalculatorRef, UBAcalculator}. The professional footprint is clearly dominated by work in a large LHC experiment, followed by travel. Work on a smaller HEP experiment using a green electricity supply or an astronomy experiment would have resulted in a smaller experimental footprint, leaving the travel footprint dominant. Several cross-continental business travels per year also would result in a significantly larger travel footprint, turning it into the leading contribution. A high or extremely high usage computing level would significantly increase the computing footprint, making it instead the leading contribution. Similarly, employment at a research centre with larger laboratory facilities could result in the institutional footprint becoming dominant. 

For the benchmark doctoral researcher and several other scenarios, the professional footprint exceeds the personal footprint, both of which by far exceed the remaining carbon budget per person per year of \SI{1.7}{\tcotwoeq} (\SI{4.8}{\tcotwoeq}) for a maximum temperature increase of \SI{1.5}{^\circ C} (\SI{2.0}{^\circ C}), as discussed in Section~\ref{introduction} and displayed in Figure~\ref{fig:Totalfootprint}. While the calculations in this study rely on several assumptions -- such as average energy consumption obtained from a few institutions and experiments, anecdotal travel habits and infrastructure usage -- that come with inherent uncertainties, it is clear that the way research is performed needs to be transformed towards a more sustainable and environmentally friendly practice. The Kyf calculator provides a first idea for an individual researcher where these transformations have to happen first and most urgently, such that the biggest and easiest contributions are addressed first, immediately followed by tackling the more challenging ones. 
Environmental reports by universities, research centres, experiments, computing centres, etc. provide key information for this purpose, and should become mandatory as soon as possible, followed by action (plans) for the reduction of the carbon footprint. This should also allow to drop or validate many assumptions which had to be made in this study.

Changes towards a more sustainable research practice will require resources, both financially and in terms of personnel. In terms of individual solutions, researchers can begin by calling for the use renewable energy in research institutes where available, and adopting more sustainable computing practices, such as optimising data processing tasks and supporting the transition to energy-efficient servers. Related to travel, prioritising virtual conferences, minimising flights, and choosing train travel where possible are essential actions. In experimental collaborations and especially in the design of future experiments, sustainability considerations and efforts need to become a central part of research activities. Additionally, a broader political and institutional effort is necessary to achieve substantial change. Together with making environmental reporting mandatory, funding bodies and policy makers must recognise the importance of sustainable research practices and allocate financial and human resources to support this transition. Public advocacy for sustainability in research should be strengthened, with calls for policies that incentivise greener technologies, carbon-neutral research infrastructures and carbon-positive technologies.
The costs of inaction will ultimately be far higher than those required for coordinated and targeted action now.

\section{Discussion}

As the climate crisis is intensifying, targeted action to reduce carbon emissions is urgently required in all fields of human activity, including research.
\emph{Knowing your footprint} is an essential first step in tackling climate change and for motivating sustainable behaviour. This holds both in personal and in professional life. 
In energy-intensive research fields, such as High Energy Physics (HEP) and related areas, which aim at the understanding of the fundamental laws of our universe, awareness is key to start moving towards more sustainable research practices. 

A growing effort to improve sustainability in HEP has largely focussed on the institutional level in terms of carbon accounting, while awareness is improved especially if numbers can be personalised.
As members of the \emph{young High Energy Physicists} (yHEP) association and as essential part of yHEP's \emph{Know your footprint} (Kyf) campaign, we have evaluated the \cotwo-equivalent emissions generated by HEP-related research on the per-researcher level. The estimate is split into the four categories: \emph{Experiments}, \emph{Institutes}, \emph{Computing} and \emph{Travel}. A user-oriented tool is provided~\cite{yHEPwebpage} to allow individual researchers in, or associated to, Germany to evaluate their professional footprint in HEP-related research. A benchmark scenario shows that for various combinations, the professional footprint significantly exceeds the personal footprint of an average citizen in Germany, which already exceeds the maximal remaining carbon budget per person per year, allowed for a maximum temperature increase of \SI{1.5}{^\circ C} (\SI{2.0}{^\circ C}), by factors of around \SI{6}{} (around \SI{2}{}). Targeted and prompt action is therefore essential.

While it is encouraging to see a growing awareness among individuals and institutions of their carbon footprint, our study shows that significant work in the HEP-related research lies ahead of us. Given that the next 10--20 years will be crucial in mitigating global warming, this work needs to start now. Carbon-footprint accounting provides us with the data to shape strategies and target measures in order to effectively reduce our carbon footprint. Scientific research needs to ensure that our pursuit of scientific and technological progress aligns harmoniously with a strong commitment to environmental sustainability.
Based on our study, experimental, institutional, computing and travel-related contributions, all need to be reduced. This requires detailed investigations of the carbon emissions of current experiments, sustainability-aware design of future experimental facilities, and a push to renewable energies for experiments, institutes, and computing centres alike. Sustainable procurement practices, improved power usage effectiveness of computing centres, efficient and resource-optimised software and prioritisation of any travels by train instead of flights are mandatory as well. The transition of the scientific system to one which places sustainability at the core of its activities, is a strong statement that we finally take the scientific results of our colleagues from climate research seriously. Every gram of \cotwo saved will make a difference.

\section{Methods}
\label{sec:methods}

The methodologies used for the professional footprint in the four distinct categories: Experiments, Institutions, Computing, and Travel, are discussed in Sections~\ref{sec:exp} to \ref{sec:travel}.

\subsection{Experiment, collaboration or project footprint}
\label{sec:exp}

The experiment, collaboration or project footprint strongly depends on the corresponding experiment, collaboration or project. In order to provide a rough estimate, we use the experiments at the Large Hadron Collider (LHC) as benchmark for large HEP experiments, DESY electricity consumption (excluding the European XFEL) as benchmark for smaller HEP experiments and the European Southern Observatory (ESO) as benchmark for experimental research in astronomy. 

To evaluate the per-researcher environmental footprint, two numbers are needed: an estimate of the total amount of emissions by the experiment, collaboration or project in tonne-\cotwo-equivalent (\tcotwoeq), and the number of people that the footprint should be distributed over. The ratio of these two numbers provides the per-person benchmark footprint.

The design of experimental setups in HEP has traditionally been driven, mainly, by the scientific question that is to be answered, without systematic consideration for minimising its carbon footprint. The scientific output is credited to the authors of the scientific publications. In the area of HEP, authorship on the publications typically requires membership in the experimental collaboration. We therefore assign the carbon footprint of the experiment to the members of the experimental collaboration, or the users (and operators) of the experimental facility, where corresponding numbers are available.

More indirect beneficiaries of the knowledge gained in the experiment, collaboration, or project, such as \emph{the industry} or \emph{the public}, are not considered here. These categories are too vague to provide a good reference for the carbon footprint of an experimental collaboration, and do not allow for well-defined boundaries of the organisation, i.e. the experiment, collaboration or project. In contrast to a company, where the footprint could alternatively be assigned to the ``product'' of the company and could thus be attributed to the  ``consumer'', \emph{knowledge} as a ``product'' of scientific experiments is not easily quantifiable and thus not distributable to ``consumers''. In order to maintain the responsibility for the emissions with those who can impact their reduction, we estimate the per-researcher footprint without consideration of such indirect benefits.
This approach is not intended to scrutinise scientists for the environmental impact of their research, but rather aims to highlight key areas where the scientific community can focus its sustainability efforts, ensuring a more environmentally responsible approach to advancing knowledge.

\subsubsection{LHC experiments}
\label{LHC}

The LHC~\cite{LHCAcc} is the largest particle accelerator housed at CERN~\cite{CERNHome} colliding proton beams at the location of four major particle detectors: ATLAS~\cite{ATLASDet}, CMS~\cite{CMSDet}, ALICE~\cite{ALICEDet} and LHCb~\cite{LHCbDet}. While ATLAS and CMS are large general-purpose detectors, ALICE and LHCb are specialised for specific investigations. Due to different sizes of the detectors, it is expected that the larger experiments also have higher annual \cotwo emissions compared to the two smaller experiments. Hence, the Kyf calculator aims at estimating separate values for annual emissions per person for the larger (ATLAS and CMS) and smaller (ALICE and LHCb) LHC experiments.

Since 2017, CERN publishes a biennial environmental report listing its \cotwo emissions. In order to estimate the annual footprints for this benchmark scenario, the environmental reports for 2017--18~\cite{CERN1718}, 2019--20~\cite{CERN1920} and 2021--22~\cite{CERN2122} were used to get an estimate of years with, as well as without, the LHC in operation. In addition, the latest technical design report from the LHCb collaboration, which includes a section on environment protection and safety, was used to estimate the difference in \cotwo emissions between the larger and smaller LHC experiments~\cite{LHCbTDR}. In this section, the annual emissions are distributed into two phases: Run phase (2017, 2018 and 2022) and Shutdown phase (2019--21).

CERN categorises its \cotwo emissions into three scopes: scope~1 for direct emissions from sources such as detector operations and heating, scope~2 for indirect emissions primarily arising from electricity consumption, and scope~3 for emissions from other indirect sources such as travel, commute, waste, catering and procurement. For the estimation of the footprint for affiliation to the LHC experiments, only scopes~1 and 2 are relevant with the following corrections applied:
\begin{itemize}
    \item Scope~1: CERN includes the emissions from heating, non-LHC experiments and others into scope~1.
    Since this estimate is only for LHC experiments, the corrected scope~1 value, corresponding to LHC experiments (particle detection and cooling) are extracted using \texttt{PlotDigitizer}~\cite{Plotdigitzer} from the figure: \emph{CERN scope~1 emissions for 2017--2022 by category}, on page 18 in CERN's environmental report 2021--2022~\cite{CERN2122}. The figure is reproduced in the supplementary material online. The total and corrected scope~1 values for all years are listed in Table~\ref{tab:LHCscope1}.
    \item Scope~2: The scope~2 contribution of CERN's \cotwo emissions include the electricity consumption of the Meyrin and Prevessin sites, which do not correspond to the consumption for powering the LHC and should be corrected for. CERN's environmental report 2021--2022 mentions that powering the CERN campus corresponds to 5\% of the total energy consumption which was used as the correction factor for the scope~2 values~\cite{CERN2122}. Since, the scope~2 emissions for 2017--20 were recalculated for the 2021--22 environment report, but only provided in a figure, they were extracted using \texttt{PlotDigitizer}~\cite{Plotdigitzer} from the figure: \emph{CERN scope~2 emissions for 2017--2022}, on page 19 in CERN's environmental report 2021--2022~\cite{CERN2122}. The values for 2021 and 2022 were directly listed in the same report. The total and corrected scope~2 values are listed in Table~\ref{tab:LHCscope2}.
\end{itemize}

\begin{table}[hbt]
\centering
\begin{subtable}{0.49\textwidth}
\centering
\begin{tabular}{c l r r}
\toprule
    & \textbf{Year}    & \multicolumn{1}{c}{\textbf{Total}}    & \multicolumn{1}{c}{\textbf{Corrected}}     \\
\midrule
\multirow{4}{*}{\rotatebox[origin=c]{90}{\textbf{Run}}} %
    & 2017  & \SI{193600}{}  & \SI{168293}{}   \\
    & 2018  & \SI{192100}{}  & \SI{162718}{}   \\
    & 2022  & \SI{184173}{}  & \SI{152444}{}   \\
    \cmidrule{2-4}
    & \textbf{Mean}  & - & \SI{161151}{} \\
\midrule
\multirow{4}{*}{\rotatebox[origin=c]{90}{\textbf{Shutdown}}} %
    & 2019  & \SI{78169}{}   & \SI{56446}{}   \\
    & 2020  & \SI{98997}{}   & \SI{75958}{}   \\
    & 2021  & \SI{123174}{}  & \SI{89547}{}   \\
    \cmidrule{2-4}
    & \textbf{Mean}  & - & \SI{73984}{} \\
\bottomrule
\end{tabular}
\caption{Scope~1}
\label{tab:LHCscope1}
\end{subtable}
\begin{subtable}{0.49\textwidth}
\centering
\begin{tabular}{c l r r}
\toprule
    & \textbf{Year}    & \multicolumn{1}{c}{\textbf{Total}}  & \multicolumn{1}{c}{\textbf{Corrected}}    \\
\midrule
\multirow{4}{*}{\rotatebox[origin=c]{90}{\textbf{Run}}} %
    & 2017  & \SI{66667}{}   & \SI{63333}{}   \\
    & 2018  & \SI{74884}{}   & \SI{71140}{}   \\
    & 2022  & \SI{63161}{}   & \SI{60003}{}   \\
    \cmidrule{2-4}
    & \textbf{Mean}  & - & \SI{64825}{} \\
\midrule
\multirow{4}{*}{\rotatebox[origin=c]{90}{\textbf{Shutdown}}} %
    & 2019  & \SI{28527}{}   & \SI{27101}{}   \\
    & 2020  & \SI{26202}{}   & \SI{24891}{}    \\
    & 2021  & \SI{56382}{}   & \SI{53563}{}   \\
    \cmidrule{2-4}
    & \textbf{Mean}  & - & \SI{35185}{} \\
\bottomrule
\end{tabular}
\caption{Scope~2}
\label{tab:LHCscope2}
\end{subtable}
\caption{Total and corrected (a) scope~1 and (b) scope~2 contribution to CERN's total emissions for the various years in Run and Shutdown phases~\cite{CERN1718,CERN1920,CERN2122}. The corrected scope-1 values were extracted using \texttt{PlotDigitizer}~\cite{Plotdigitzer} from the figure: \emph{CERN scope~1 emissions for 2017--2022 by category}, on page 18 in CERN's environmental report 2021--2022~\cite{CERN2122}. Scope-2 values for 2017--20 were extracted using \texttt{PlotDigitizer}~\cite{Plotdigitzer} from the figure: \emph{CERN scope~2 emissions for 2017--2022}, on page 19 in CERN's environmental report 2021--2022~\cite{CERN2122}. The corrected scope-2 value is 95\% of the total value. All emission values are provided in \SI{}{t \cotwo e}.}
\label{tab:LHCscopes}
\end{table}

The technical design report of the LHCb experiment lists their total emission for 2022 to be \SI{4400}{t \cotwo e} of which 51\% is assigned to scope~1~\cite{LHCbTDR}, i.e. 
\begin{align*}
    \text{S1}_\text{Small}^\text{Run} &= 0.51 \times \SI{4400}{t \cotwo e} = \SI{2244}{t \cotwo e}\,.
\end{align*}
Since the LHC was operating in 2022, $\text{S1}_\text{Small}^\text{Run}$ is defined as the scope~1 contribution for the smaller experiments in the Run phase. In order to estimate the scope~1 contribution for the smaller experiments in the Shutdown phase ($\text{S1}_\text{Small}^\text{SD}$), it is assumed that the emissions for larger and smaller experiments scale by the same factor between the two phases. This factor ($\text{S1}^\text{Run/SD}$) can be calculated from the mean corrected scope~1 contribution for the two phases from Table~\ref{tab:LHCscope1} as
\begin{align*}
    \text{S1}^\text{Run/SD} &= \frac{\SI{161151}{}}{\SI{73984}{}} = 2.18\,.
\end{align*}
This results in $\text{S1}_\text{Small}^\text{SD}$ being estimated as
\begin{align*}
    \text{S1}_\text{Small}^\text{SD} &= \frac{\text{S1}_\text{Small}^\text{Run}}{\text{S1}^\text{Run/SD}} = \frac{\SI{2244}{t \cotwo e}}{2.18} = \SI{1030}{t \cotwo e}\,.
\end{align*}

The scope~1 contribution for the larger experiments ($\text{S1}_\text{Large}$) for each phase can be estimated using the mean corrected scope~1 contribution for all LHC experiments ($\text{S1}_\text{All}$) from Table~\ref{tab:LHCscope1} and $\text{S1}_\text{Small}$. This is done by assuming an equal contribution from both the smaller experiments and an equal contribution also from the two larger experiments and calculated as
\begin{equation}
    \text{S1}_\text{All} = 2\times\text{S1}_\text{Small} + 2\times\text{S1}_\text{Large}
    \quad\Rightarrow\quad
    \text{S1}_\text{Large} = \frac{\text{S1}_\text{All} - 2\times \text{S1}_\text{Small}}{2}\,.
\end{equation}
The resulting values for the larger experiments are
\begin{align*}
    \text{S1}_\text{Large}^\text{Run} &= \SI{78332}{t \cotwo e} \quad \text{and} \quad \text{S1}_\text{Large}^\text{SD} = \SI{35962}{t \cotwo e}\,.
\end{align*}

To estimate the scope~2 contribution of individual experiments, it is assumed that all four experiments share an equal contribution to the mean corrected scope~2 emissions of all experiments ($\text{S2}_\text{All}$) given in Table~\ref{tab:LHCscope2}. This assumption is based on the premise that the dominant scope~2 emissions come from the electricity consumption for running the entire accelerator complex, up to and including the LHC, which is equally necessary for all four experiments. In reality, the electricity consumed by the four experiments might vary and could be refined in the future if more data becomes available. The scope~2 contributions for the smaller ($\text{S2}_\text{Small}$) and larger ($\text{S2}_\text{Larger}$) experiments, calculated individually for the Run and Shutdown phases, are given as
\begin{align}
    \text{S2}_\text{Small} = \text{S2}_\text{Large} = \frac{\text{S2}_\text{All}}{4}\,.
\end{align}
Using this relation, the scope~2 contributions for the Run and Shutdown phases can be calculated as
\begin{align*}
    \text{S2}_\text{Small}^\text{Run} = \text{S2}_\text{Large}^\text{Run} &= \frac{\SI{64825}{t \cotwo e}}{4} = \SI{16206}{t \cotwo e}\,,
\end{align*}
and
\begin{align*}
    \text{S2}_\text{Small}^\text{SD} = \text{S2}_\text{Large}^\text{SD} &= \frac{\SI{35185}{t \cotwo e}}{4} = \SI{8796}{t \cotwo e}\,.
\end{align*}

After summing the contributions from scopes~1 and 2 to obtain the total footprint of smaller and larger experiments in the Run and Shutdown phases, a weighted mean of the two phases is taken assuming a Run phase of \SI{4}{years} and Shutdown phase of \SI{3}{years},
 \begin{equation}
    \label{eq:PhaseOverall}
    \text{Overall annual emission} = \frac{4\times\text{Total}^\text{Run} + 3\times\text{Total}^\text{SD}}{7}\,.
 \end{equation}
 
The values for the individual scopes in the Run and Shutdown phases for smaller and larger experiments are summarised in Table~\ref{tab:LHCexp} along with the total values for each phase and the overall values calculated using Equation~\ref{eq:PhaseOverall}.

\begin{table}[hbt]
\centering
\begin{tabular}{c l r r r}
\toprule
    & \textbf{Phase}    & \multicolumn{1}{c}{\textbf{Scope~1}}    & \multicolumn{1}{c}{\textbf{Scope~2}}  & \multicolumn{1}{c}{\textbf{Total}}    \\
\midrule
\multirow{3}{*}{\rotatebox[origin=c]{90}{\textbf{Small}}} %
    & Run   & \SI{2244}{}    & \SI{16206}{}   & \SI{18450}{}   \\
    & SD    & \SI{1030}{}    & \SI{8796}{}    & \SI{9826}{}    \\
    \cmidrule{2-5}
    & \textbf{Overall}  & - & - & \SI{14754}{} \\
\midrule
\multirow{3}{*}{\rotatebox[origin=c]{90}{\textbf{Large}}} %
    & Run   & \SI{78332}{}   & \SI{16206}{}   & \SI{94538}{}   \\
    & SD    & \SI{35962}{}   & \SI{8796}{}    & \SI{44758}{}   \\
    \cmidrule{2-5}
    & \textbf{Overall}  & - & - & \SI{73204}{} \\
\bottomrule
\end{tabular}
\caption{Contribution from individual scopes in the Run and Shutdown phases for smaller and larger experiments along with the total values for each phase and the overall values. The total values for each phase are calculated as $\text{Total} = \text{Scope~1} + \text{Scope~2}$, and the overall values are calculated using Equation~\ref{eq:PhaseOverall}.}
\label{tab:LHCexp}
\end{table}

Finally, to calculate the per person annual footprint for someone affiliated with one of the smaller (larger) LHC experiments, the overall contribution from Table~\ref{tab:LHCexp} is divided by the mean of total members listed on the public webpages of ALICE~\cite{ALICEmemb} and LHCb~\cite{LHCbmemb} (CMS~\cite{CMSmemb} and ATLAS~\cite{ATLASmemb}) experiments. These values are listed in Table~\ref{tab:LHCmembers}. The overlap of members between the four experiments is assumed to be negligible. The annual emission per person for smaller (larger) LHC experiments are estimated to be \SI{8.76}{t \cotwo e} (\SI{11.91}{t \cotwo e}).

\begin{table}[hbt]
\centering
\renewcommand{\arraystretch}{1.10}
\begin{tabular}{c l c c r}
\toprule
    & \textbf{Experiment}    & \textbf{Members}    & \textbf{Mean}    & \multicolumn{1}{c}{\textbf{Emissions}}    \\
\midrule
\multirow{2}{*}{\rotatebox[origin=c]{90}{\textbf{Small}}} %
    & ALICE & \SI{1968}{}~\cite{ALICEmemb}   & \multirow{2}{*}{\SI{1684}{}}  & \multirow{2}{*}{\SI{8.76}{t \cotwo e}}  \\
    & LHCb  & \SI{1400}{}~\cite{LHCbmemb}   & & \\
\midrule
\multirow{2}{*}{\rotatebox[origin=c]{90}{\textbf{Large}}} %
    & CMS   & \SI{6288}{}~\cite{CMSmemb}   & \multirow{2}{*}{\SI{6144}{}}  & \multirow{2}{*}{\SI{11.91}{t \cotwo e}} \\
    & ATLAS & \SI{6000}{}~\cite{ATLASmemb}   & & \\
\bottomrule
\end{tabular}
\caption{Total members affiliated to the four LHC experiments. The values in the \emph{Mean} column are calculated independently for the two smaller and two larger LHC experiments.  The \emph{Emissions} column lists the per person annual emission for the smaller and larger LHC experiments. This is calculated by dividing the \emph{Overall} emissions from Table~\ref{tab:LHCexp} by the respective values in the \emph{Mean} column.}
\label{tab:LHCmembers}
\end{table}

\subsubsection{DESY}
\label{DESY}

The research centre \emph{Deutsches Elektronensynchrotron} (DESY) in Hamburg, Germany, is the German national laboratory for accelerator development and operation, photon science, particle physics and astroparticle physics. Due to the presence of a significant accelerator and experimental complex, emissions from DESY are used as proxy for emissions from smaller HEP experiments.

The first DESY sustainability report for 2019--2021 \cite{DESY1921} has been published in 2022, but does not include detailed numbers on the emissions from scope 1, 2 or 3 sources as provided in the CERN environmental reports~\cite{CERN1718,CERN1920,CERN2122}. The DESY webpage, however, provides a figure on the energy consumption at DESY in 2021~\cite{DESYenergy} which is reproduced in the supplementary material online for long-term reference.

Electricity consumption in the year 2021 at DESY, excluding the electricity needed for the European XFEL, amount to \SI{152.5}{GWh}. This includes: the accelerator complex with the accelerators: PETRA, FLASH, LINAC, HERA and DESY (55\%), cryogenics (25\%), as well as offices, workshops and laboratories, computing centre, cooling centre, canteen and others (20\%).
Subtracting the contribution from offices, workshops and laboratories, canteen and others as institute footprint, and the computing centre as computing footprint, which are treated separately in the Kyf calculator, the remaining experimental electricity consumption is \SI{128.3}{GWh} for 2021, which is considered as representative for the annual consumption.

Two conversion factors for electricity consumption to \cotwo emissions are employed, based on data of Ref.~\cite{ElectricityMaps}: The specific \cotwo emissions averaged over the German grid in 2023 correspond to \SI{416}{g\cotwo e/kWh}; emissions from green energy production assume \SI{100}{\%} photovoltaic (PV)-based electricity production with a footprint of \SI{35}{g\cotwo e/kWh}. Emissions from wind- (\SI{13}{g\cotwo e/kWh}) and water- (\SI{11}{g\cotwo e/kWh}) based production are lower, so \SI{100}{\%} PV-based production is considered to be a conservative estimate for green electricity production.

The number of guest scientists using the DESY facilities is listed as 3000 in the DESY environmental report~\cite{DESY1921}, which is complemented by an additional 200 accelerator operators. The latter is based on an internal estimate by one accelerator operator, and is considered reasonable given the size of the DESY facilities. Refinements of these numbers are possible in a future version of the Kyf calculator.

Based on the above conversion factors and the number of scientists and operators, the total emissions from DESY are \SI{53372.8}{\tcotwoeq} (\SI{4490.5}{\tcotwoeq}) for conventional (green) electricity production, corresponding to \SI{16.68}{\tcotwoeq} (\SI{1.40}{\tcotwoeq}) emissions per person per year, given a conventional (green) electricity supply. 
DESY itself switched to a green electricity contract at the start of 2023, significantly reducing its environmental footprint.

\subsubsection{European Southern Observatory}
\label{ESO}

Since 2020, the annual reports for ESO include a section on the environmental impact of the observatory~\cite{ESO2020,ESO2021,ESO2022}. In order to estimate the annual \cotwo emissions per person, only the emissions for 2021 were used to stay with the most up-to-date values.
This, however, involves contributions from business travels, commute and waste, which do not fit well under this category and need to be subtracted. The breakdown for the various categories was extracted using \texttt{PlotDigitizer}~\cite{Plotdigitzer} from the figure: \emph{ESO \cotwo emissions 2018--2021}, on page 111 in ESO's annual report 2022~\cite{ESO2022} and is shown in Table~\ref{tab:ESOsplit}. The overall emission after correcting for the contributions from business travels, commute and waste is estimated to be $\SI{19363}{\tcotwoeq}$.

\begin{table}[hbt]
\centering
\begin{tabular}{l r}
\toprule
\textbf{Category}   & \multicolumn{1}{c}{\textbf{Emission}}    \\
\midrule
Energy (E)          & \SI{8599}{}    \\
Purchase (P)        & \SI{5924}{}    \\
Freight (F)         & \SI{2675}{}    \\
Travel (T)          & \SI{191}{}     \\
Commuting (C)       & \SI{1083}{}    \\
Capital Goods (CG)  & \SI{2166}{}    \\
Waste (W)           & \SI{127}{}    \\
\midrule
Total               & \SI{20764}{}   \\
\midrule
Corrected           & \SI{19363}{}   \\
\bottomrule
\end{tabular}
\caption{Breakdown of the total \cotwo emissions of ESO for 2021. The contribution of the various categories are extracted using \texttt{PlotDigitizer}~\cite{Plotdigitzer} from the figure: \emph{ESO \cotwo emissions 2018--2021}, on page 111 in ESO's annual report 2022~\cite{ESO2022}. The corrected value is calculated as \mbox{$\text{Corrected} = \text{Total} - \text{T} - \text{C} - \text{W}$}. All emission values are provided in \SI{}{t \cotwo e}.}
\label{tab:ESOsplit}
\end{table}

The public webpage of ESO lists more than \SI{22000}{} users of the observatory who share this overall emission~\cite{ESOpubpage}. The average annual emission per person is hence estimated as \mbox{$\SI{19363}{\tcotwoeq} / \SI{22000}{} = \SI{0.88}{\tcotwoeq}$}.

More refined evaluation of individual ESO facilities and experiments would be very interesting for future versions of the Kyf calculator, if additional information becomes available.

\subsection{Institute or research centre footprint}
\label{sec:inst}

Institutes within universities or research centres, as key academic institutions, are an integral part of a researcher's environmental footprint. As benchmarks, the University of Freiburg and CERN are considered in the Kyf calculator as proxies for a German university and a research centre, respectively. The choice is based on the location of the university in Germany and its involvement in HEP research, and CERN as central European HEP laboratory, as well as the publication of environmental reports by the institutions with sufficient information for a calculation. Among other universities with environmental reports such as the Bauhaus-University Weimar~\cite{UniWeimar}, the University Tübingen~\cite{UniTuebingen}, and the Leibniz University Hannover~\cite{UniHann}, the latter has been investigated as a cross-check for a German university's footprint, with details provided in the supplementary material online. In the three cases, procurement which is important for the environmental report of the University of Freiburg is either partially or not considered. In the future, more environmental reports by universities are highly encouraged and welcome, in particular with joint accounting metrics.

Our analysis is based on the environmental reports, covering the years 2019--2020 for the University of Freiburg~\cite{UniFR}, and the years 2021--2022 for CERN~\cite{CERN2122}. We base our estimate of the institutional footprint on one example year each (outside of the COVID-19 pandemic), and consider this to be representative for the annual footprint. Future assessments will test this assumption.

The carbon footprint of the institutions is assessed across several categories including electricity, heating/cooling, water, waste, and procurement. The section offers a detailed comparison between these entities, highlighting the diversity and scale of their emissions. Special attention is given to the significant role of procurement in contributing to the overall carbon footprint of universities and research centres. 

\subsubsection{University of Freiburg}
\label{UniFreiburg}

The environmental footprint for the University of Freiburg is based on the environmental report for 2019/2020~\cite{UniFR}, published in 2021. The numbers for 2019 are used as the most recent, pre-pandemic numbers considered to be representative for the annual consumption of the university.

Emissions of the University of Freiburg are included for the categories: Electricity, heating/cooling, water, waste, and procurement. Numbers for vehicle fleet and business travel are provided as well, but are considered separately in the Kyf calculator and are therefore not listed here. The numbers provided for each of the used categories are reproduced in Table~\ref{tab:UniFRsplit}. Additional detail on the numbers and categories for procurement is provided in the supplementary material online.

\begin{table}[hbt]
\centering
\begin{tabular}{l r}
\toprule
\textbf{Category}   & \multicolumn{1}{c}{\textbf{Emissions [\tcotwoeq]}}    \\
\midrule
Electricity     & \SI{2431}{} (\SI{19224}{})   \\
Heating/Cooling & \SI{13584}{}    \\
Water           & \SI{14}{}    \\
Waste           & \SI{577}{}     \\
Procurement     & \SI{14486}{}    \\
\midrule
Total           & \SI{31092}{} (\SI{47885}{}) \\
\bottomrule
\end{tabular}
\caption{Breakdown of the total \cotwo emissions of the University of Freiburg for 2019~\cite{UniFR}. All categories are summed except \emph{Vehicle fleet + business travel} which is considered separately in the Kyf calculator. Electricity values are provided nominally using green electricity supply. The values in brackets provide the numbers using electricity produced with the German electricity mix. The conversions to \tcotwoeq are conducted by the University of Freiburg.}
\label{tab:UniFRsplit}
\end{table}

The total annual footprint of the University of Freiburg, excluding vehicle fleet and business travel which are considered separately in the Kyf calculator, is estimated as \SI{31092}{\tcotwoeq} (\SI{47885}{\tcotwoeq}) using green electricity (the German electricity grid mix). The University of Freiburg obtains green electricity from certified water-power plants~\cite{UniFR}. The largest individual contributor for the University of Freiburg, when considering green electricity supply, is procurement which corresponds to \SI{87}{\%} of the emissions from other categories (excl. vehicle fleet + business travel).

The environmental footprint of the University of Freiburg is distributed over the members of the university (students and employees) which are determined as \SI{31147}{} from the environmental report~\cite{UniFR}. The institutional footprint per person per year therefore amounts to \SI{1.00}{\tcotwoeq} (\SI{1.54}{\tcotwoeq}) using green electricity (the German electricity grid mix).

\subsubsection{Research centre CERN}
\label{CERN}

The environmental footprint for CERN as research centre is based on the values for 2022 (post-pandemic), listed in the newest environmental report for 2021--2022, published in 2023~\cite{CERN2122}.

The categories considered are similar to the ones for University of Freiburg: \emph{Electricity}, \emph{Heating (gas+fuel)} and \emph{Other (scope 1)}, \emph{Water purification}, \emph{Waste}, \emph{Procurement}. A value for business travel is provided as well, but will be considered separately in the professional footprint of the Kyf calculator and is therefore dropped here. Values for catering and personal commutes are provided by the CERN environmental report, in addition; however, in the Kyf calculator, food, diet choices and commute are considered as part of the personal footprint, and are therefore removed from the institutional footprint to avoid double counting. 
The numbers for each of the employed categories are provided in Table~\ref{tab:CERNsplit}.

\begin{table}[hbt]
\centering
\begin{tabular}{l r}
\toprule
\textbf{Category}   & \multicolumn{1}{c}{\textbf{Emissions [\tcotwoeq]}}    \\
\midrule
Electricity     & \SI{3158}{}   \\
Heating (gas+fuel) + Other & \SI{11250}{}    \\
Water purification         & \SI{176}{}    \\
Waste           & \SI{1875}{}     \\
Procurement     & \SI{104974}{}    \\
\midrule
Total           & \SI{121433}{} \\
Total without Procurement & \SI{16459}{} \\
\bottomrule
\end{tabular}
\caption{Breakdown of the total \cotwo emissions of CERN as research institute for 2022~\cite{CERN2122}. The categories \emph{Business travel} and \emph{Personnel commutes} are considered separately in the Kyf calculator, and thus not listed here. The conversions to \tcotwoeq are conducted by CERN.}
\label{tab:CERNsplit}
\end{table}

Emissions for electricity correspond to \SI{5}{\%} of the total electricity-related emissions for 2022, which is specified in the CERN environmental report 2021--2022~\cite{CERN2122} as electricity component required for powering the campus. Though not explicitly specified in the report, the effective conversion factor for electricity consumption to \cotwo emissions can be derived as \SI{52}{\tcotwoeq/GWh (=g\cotwo e/kWh)} from the total electricity consumption in 2022 of \SI{1215}{GWh}, corresponding to emissions of \SI{63161}{\tcotwoeq}, specified for scope 2 in 2022. Electricity at CERN largely comes from low-carbon nuclear-power plants from France, resulting in a conversion factor significantly lower than the one from the conventional German grid mixture, but still slightly larger than electricity based on renewables, discussed in Section~\ref{DESY}. The numbers for \emph{Heating (gas+fuel) and Other} (scope 1) are obtained from the total scope-1 emissions minus emissions for \emph{LHC experiments: Particle detection} and \emph{LHC experiments: Detector cooling} as well as \emph{Other experiments}, where the LHC-related emissions are considered as part of the experimental contribution, discussed in Section~\ref{LHC}. The value for the sum of the LHC and non-LHC experimental contributions has been obtained using \texttt{PlotDigitizer}~\cite{Plotdigitzer} (also see the supplementary material online). Additional detail on the composition of the procurement category are given in the supplementary material online.

The total emissions for CERN as a research centre in 2022 sum to \SI{121433}{\tcotwoeq} (\SI{16459}{\tcotwoeq}) including (excluding) procurement. 
They are dominated by the footprint of procurement, followed by heating (gas+fuel), where heating reaches a similar level as for the University of Freiburg. A split up of the emissions from procurement is provided by the CERN environmental report 2021--2022~\cite{CERN2122}, and discussed further in the supplementary material online. It is important to note that the emissions from procurement at CERN do not only cover those related to CERN as an institute, but also those related to the experiments, accelerator facilities, as well as future accelerator developments. 
Some procurement categories such as \emph{Civil engineering, building and technical services} appear more related to the infrastructure of the accelerator and experimental complex; different categories such as \emph{Other} that includes office supplies, furniture, etc. seem more related to CERN as an institute. A clear separation of the procurement categories belonging to CERN as an institute is however not possible. In the future, improved assignment can be done if more granular information becomes available. For the moment, procurement is fully assigned to the CERN institute footprint. 

Emissions from electricity and waste play a smaller role in the carbon footprint of CERN as an institute. Differences in the electricity consumption to the university footprint can be primarily attributed to different conversion factors for the type of used electricity. The actual electricity consumption of the CERN campus in GWh is \SI{60.75}{GWh}, corresponding to \SI{5}{\%} of the total consumption of \SI{1215}{GWh} in 2022, which is comparable to that of the University of Freiburg and the Leibniz University Hannover, with \SI{50}{GWh} and \SI{58.66}{GWh}, respectively. 

The evaluation of an \emph{effective CERN population} needs to consider that at any time during the year, a certain fraction of CERN users is present at CERN, using electricity, water, heating, etc. and thus contributing to the institutional footprint. The total number of CERN personnel employed by CERN (Employed Members of Personnel - MPE) in 2022 amounts to $N_\text{MPE}=\SI{3558}{}$~\cite{CERNpop}, including staff and CERN fellows. The total number of Associated Members of Personnel (MPA), including CERN users as largest subcategory, corresponds to $N_\text{MPA}=\SI{13376}{}$ in 2022~\cite{CERNpop}. The average presence for MPAs at CERN is evaluated as a weighted average of the \emph{Yearly Presence} of CERN users at CERN, using the midpoint of the percentage bins in Table 11 of Ref.~\cite{CERNpop} for weighting. This results in an average presence $\bar{p}=\SI{27.9}{\%}$ for MPAs at CERN. The effective CERN population, $N_\text{eff}$, is thus calculated as:
$$N_\text{eff} = N_\text{MPE}+\bar{p} \cdot N_\text{MPA}\,,$$
and, numerically, determined to be $N_\text{eff}=\SI{7295}{}$.

The institutional footprint per person per year for CERN as research institute is estimated to be \SI{16.65}{\tcotwoeq} (\SI{2.26}{\tcotwoeq}) including (excluding) procurement. 
This is, also excluding procurement, larger than the institute footprint of the University of Freiburg, discussed in Section~\ref{UniFreiburg}. Given the larger experimental facilities at the research centre, higher per-researcher emissions at a research centre compared to a university are expected.
The impact of procurement at CERN as an institute is artificially increased, since it also includes contributions relevant to the experiments, accelerators and future developments, which cannot be separated at the moment. 
The estimate of emissions from procurement at CERN are based on the Greenhouse Gas Protocol spend-based method, which -- though already rather advanced -- is acknowledged in the CERN environmental report 2021-2022~\cite{CERN2122} to have limitations and variable degrees of uncertainty. CERN started the \emph{Environmentally Responsible Procurement Policy Project (CERP3)} in September 2021, to improve the environmental impact of procurement in the future.

\subsection{Computing footprint}
\label{sec:comp}

Most of research in physical sciences today relies on having High Performance Computing (HPC) Clusters to enable breakthroughs and make advancements within the various sub-fields of physics. Hence, it is essential to consider the environmental impact of such computing clusters. Sustainable research and accelerating scientific progress is very much dependent on rise of awareness of our resource usage \cite{bruers2023resourceaware}. 

Typical HPC usage involves researchers submitting jobs to computational clusters via a scheduler or interactively. For each job, specific requirements are defined, including the number of CPU-cores or GPUs needed and a defined time window for execution. For a carbon footprint, average idle times as well as surplus-power, e.g. due to HPC cooling, need to be considered as well.
The carbon footprint averaged over a year is thus estimated as: 
\begin{equation}
    \text{Total Footprint [\tcotwoeq]} = f_\text{PUE} \cdot f_\text{overh} \cdot n_\text{WPC} 
    \cdot f_\text{conv}\,,\\ 
    \label{AnnualCO2FpComp1}
\end{equation}
where $f_\text{PUE}$ denotes the HPC's \emph{Power Usage Effectiveness} (PUE) \cite{PUE}, $f_{overh}$ is an overhead factor to account for power consumption of the HPC when its computing cores are idle, $n_\text{WPC}$ accounts for the \emph{Workload Power Consumption} (WPC), and $f_\text{conv}$ is the conversion factor from kWh to \gcotwoeq.
The yearly WPC is estimated taking into account the usage of CPUs and GPUs separately as:
\begin{equation}
    n_\text{WPC} = p_\text{CPU-core}\cdot l_\text{core-h, CPU} + p_\text{GPU}\cdot l_\text{h, GPU}\,,
    \label{AnnualCO2FpComp2}
\end{equation}
where $p_\text{CPU-core}$ is the power consumption in kW for each CPU core and $l_\text{core-h, CPU}$ is the CPU workload measured in core hours. GPU usage is calculated with $p_\text{GPU}$ representing the power consumption in kW per GPU and $l_\text{h, GPU}$ is the total number of hours the GPU is used for.
We do not detail GPU power consumption by individual CUDA cores \cite{cuda} for two main reasons: the standard power specification for a GPU, discussed below, offers a reliable estimate, and the real-world application of GPUs does not involve partitioning by CUDA cores, rendering such a breakdown impractical here.

For the Kyf calculator, the following values are used as default values for an individual researcher's computing footprint:
\begin{itemize}
 \item For $f_\text{PUE}$, a value of 1.5 has been chosen, which is listed in the CERN environmental report 2021--2022~\cite{CERN2122} as global average PUE for large data centres, and which has been declared as the target to be reached by 1 July 2027 for computing centres starting operations before 1 July 2026 in the German law for the improvement of energy efficiency (\emph{Energieeffizienzgesetz} - EnEfG, §11(1))~\cite{EnEfG}. 
 \item The idle-time overhead factor $f_\text{overh}$ has been estimated by comparing the system power consumption of the \emph{Hawk} supercomputer at the HPC Stuttgart~\cite{Hawk} at full load (LinPack operation) (\SI{4.1}{MW}) to normal operation (\SI{3.5}{MW}). The resulting factor of \SI{1.17}{} is used as default idle-time overhead factor. The idle-time overhead results from the HPC having to estimate the number of cores it operates by predicting the requirements by its users. When too few cores are operated, it leads to long waiting times for jobs to start, but no idle operation time. When too many cores are available, idle times become too large. HPCs will thus aim for a small fraction of idle time, which needs to be considered in the carbon footprint.
 \item As conversion factor $f_\text{conv}$, the same values as discussed in Section~\ref{DESY} are employed, based on Ref.~\cite{ElectricityMaps}: \SI{416}{\gcotwoeq/kWh} (\SI{35}{\gcotwoeq/kWh}) for a conventional (green) electricity provision at the HPC. 
 \item The default value for the CPU power consumption per core, $p_\text{CPU-core}$, is set to \SI{7.25}{W}, and for the default power consumption per GPU,  $p_\text{GPU}$, to $\SI{250}{W}$ . The CPU value is obtained from the DESY Maxwell cluster with AMD EPYC 75F3 CPU cores \cite{DESYMaxwell_private_communication}. The GPU value is the median of a range, \SI{150}{}--\SI{350}{W}, reported on a forum of NVIDIA GPU users \cite{NVIDIA_A100_GPU}. Both these systems are fairly recent, so the power usages should be representative of newer HPC processing units. 
\end{itemize}

The usage load in core-hours for CPU and GPU usage, $l_\text{core-h, CPU}$ and $l_\text{h, GPU}$, needs to be provided by the researcher when using the Kyf calculator.
The computing footprint estimate can be further personalised by adjusting $f_\text{PUE}$, if calculations have been done, for example, at the very new CERN computing centre which targets a PUE value of 1.1~\cite{CERN2122}, or by changing $f_\text{overh}$ to a different value, if better or worse HPC load factors have been obtained. It can also be switched to green electricity, with conventional electricity being used as default.
In recent years, HPC centres have been implementing system-generated reports that inform their users about their usage with high transparency. Resources like the \emph{Production and Distributed Analysis} (PanDA) client~\cite{Big_Panda} or API~\cite{Panda_API} can also be used by users or software developers to monitor group or individual workload on a system. If the annual carbon footprint is thus known precisely, this can also be entered directly in  \tcotwoeq. 

For convenience, four scenarios discussed in the following are provided, based on the estimated usage level: low, medium, high and extremely high. Several factors come into play when deciding whether to utilise a CPU or a GPU for a particular computational task. These factors include the availability of suitable software and libraries, memory requirements, considerations related to parallelisation, as well as potential bottlenecks in data transfer rates. As a result, simulations and analyses are tailored to the specific needs and research objectives of individual researchers. The provided examples thus serve as rough approximations only, and CPU vs. GPU usage cannot necessarily be interchanged easily.

\begin{table}[hbt]
\centering
\begin{tabular}{l c c c}
\toprule
\textbf{Scenario}   & \textbf{CPU [core-h/month]} & \textbf{GPU [h/month]} & \textbf{Annual footprint [\tcotwoeq]} \\
\midrule
Low usage & \SI{4000}{} & - & 0.25 \\
Medium usage & \SI{30000}{} & - & 1.91 \\
High usage & - & 2500 & 5.48 \\
Extremely high usage & - & 8000 & 17.52 \\
\bottomrule
\end{tabular}
\caption{Annual footprint in \tcotwoeq obtained for four different scenarios of computing usage of an individual researcher. Default numbers as discussed in the text are used, including the conversion factor $f_\text{conv}$ for conventional electricity provision at the HPC.}
\label{tab:CompScenarios}
\end{table}

\begin{itemize}
\item{Case for low usage} \newline
Example: A graduate student submits several jobs per week each needing between one to four CPU cores. The consumption corresponds to an an average of \SI{4000}{\text{CPU core-h}} used per month.
\item{Case for medium usage} \newline
Example: A doctoral student or post-doctoral researcher, involved in data analysis 70-100\% of their time, submits jobs over the year. 
For this case, the top five ranked users at the Uni-Freiburg HPC called the Black-Forest Grid (BFG) are used as a blueprint, giving \SI{30 000}{CPU\text{ core-h}} per month.
\item{Case for high usage} \newline
Example: An accelerator scientist studies accelerator performance with particle tracking codes and semi particle-in-cell (PIC) codes. The studies are both memory and computationally demanding.
We assume \SI{2500}{\text{GPU h}} used per month with code optimised for GPUs (equivalent to \SI{80 000}{\text{CPU core-h}} per month).
\item{Case for extremely high usage} \newline
Example: A researcher runs PIC simulations or high-resolution imaging analysis. This corresponds to a usage of \SI{8000}{\text{GPU h}} per month (equivalent to \SI{300000}{\text{core-h}} per month on CPUs).
\end{itemize}

The default numbers as discussed above are used, including the conversion factor $f_\text{conv}$ for conventional electricity provision at the HPC, to convert the monthly CPU or GPU consumption into annual footprints. The results are listed in Table~\ref{tab:CompScenarios}.

It is important to acknowledge that the calculations presented here are based on the assumption of optimal core utilisation in HPCs. In practice, however, it is often the case that not all cores are used efficiently, leading to unnecessary energy consumption. 

The environmental costs associated with data centres for long-term storage require significant amounts of electricity and water to operate servers and cooling systems. The large scale resources used by experimental collaborations or at institutes are assumed to be accounted for within their respective categories. Within the individual computing footprint, the environmental contribution of data storage is assumed to be small compared to that of actively running jobs and is therefore neglected. If we obtain data that such costs for an individual are high, we will include this contribution in future iterations.
If significant amounts of data are stored at a location other than the institute, such as on a commercial cloud, e.g. Microsoft OneDrive~\cite{onedrive} or Google Drive~\cite{googledrive}, an additional \cotwo emission contribution can be added, based e.g. on the Microsoft emissions impact dashboard~\cite{onedrive_dashboard}, or the Google's Carbon Footprint app~\cite{googledrive_footprintcalculator}.

\subsection{Business travel footprint}
\label{sec:travel}

Travel constitutes another crucial component of High Energy Physics (HEP). The importance of in-person events became notably evident during the COVID-19 pandemic, where their absence made discussions and collaborations more challenging, prompting a reconsideration of our methodologies. In light of this realisation, a meticulous assessment of travel-associated \cotwo emissions is imperative. Business travel, including transportation and accommodations, significantly contributes to \cotwo footprints. Efforts to comprehend and mitigate these emissions involve scrutinising various transportation modes, such as long-distance buses, trains, personal cars, and air travel, each exerting diverse impacts on our environmental footprint~\cite{UBAcalculator, Umweltbundesamt_TREMOD}. Comparison of \cotwo emissions from various modes of transport are briefly discussed in the supplementary material online. Additionally, the emissions from factors such as accommodation and event venues play a role, emphasising the need to evaluate the average \cotwo emissions of hotel rooms per night and event venues per day for an individual~\cite{VeranstaltungRechnerUBA}. Benchmark values used for the Kyf calculator are documented in Table~\ref{BusinessTravelTable}. These values are estimated using travel, hotel stay and meeting room usage scenarios in Germany. The footprint corresponding to the event venue assumes an area of \SI{4}{m^2} per person.

\begin{table}[hbt]
\centering
\begin{tabular}{l r l}
\toprule
\textbf{Source of Emission} & \multicolumn{2}{c}{\textbf{Emission Factor}} \\
\midrule
Long-distance Buses         & \SI{0.031}{}      & \SI{}{kg \cotwo e / km}   \\
Long-distance Trains        & \SI{0.031}{}      & \SI{}{kg \cotwo e / km}   \\
Personal Car                & \SI{0.17}{}      & \SI{}{kg \cotwo e / km}   \\
Flights within Europe       & \SI{130}{}    & \SI{}{kg \cotwo e / h}    \\
Transcontinental Flights    & \SI{170}{}    & \SI{}{kg \cotwo e / h}    \\
Hotel room                  & \SI{12}{}      & \SI{}{kg \cotwo e / night}\\
Event venue                 & \SI{0.19}{}      & \SI{}{kg \cotwo e / day}    \\
\bottomrule
\end{tabular}
\caption{List of dominant contributors to the \cotwo footprint of a business trip. The dominant contributors are indicated together with the respective \cotwo-equivalent emission value for an average individual in Germany.}
\label{BusinessTravelTable}
\end{table}

It is important to note that some institutes, such as the University of Freiburg, compensate for all flights taken by its employees for business trips. The Kyf calculator provides an option for showing the amount of carbon footprint, which was compensated. This might be extended to other sections of the Kyf calculator in future iterations.

Using the values from Table~\ref{BusinessTravelTable}, it is possible to obtain an approximate estimate of the carbon footprint for a business trip. Three benchmark scenarios are presented here and are summarised in Table~\ref{tab:TravScenarios}.

\begin{itemize}
    \item Train travel within Europe: A researcher in Freiburg attending a conference in Hamburg lasting \SI{5}{days} from Monday to Friday and travelling by long-distance trains. One-way distance between Freiburg and Hamburg is about \SI{800}{km}. The scenario assumes a hotel stay of \SI{5}{nights} from Sunday to Friday and occupancy of the event venue for  \SI{8}{h/day} for \SI{5}{days}.
    \item Flight travel within Europe: A researcher based in Freiburg attending a conference lasting \SI{5}{days} in Thessaloniki, Greece. A direct flight from Basel to Thessaloniki takes about \SI{2.5}{hours} for one way. Assumptions made on the hotel stay and event venue occupancy are same as in the first scenario.
    \item Flight travel across continents: A researcher travelling from Freiburg to Seoul, South Korea for a summer school lasting \SI{2}{weeks}. A one-way travel requires a long-distance train from Freiburg to Frankfurt (\SI{275}{km}) followed by an transcontinental flight from Frankfurt to Seoul (\SI{12}{hours}). A hotel reservation of \SI{14}{nights} is assumed along with event venue occupancy of \SI{8}{h/day} for \SI{10}{business\ days}.
\end{itemize}

\begin{table}[hbt]
\centering
\begin{tabular}{l c c c c c}
\toprule
\textbf{Scenario} & \textbf{Train [km]} & \textbf{Flight [h]} & \textbf{Hotel} & \textbf{Venue} & \textbf{Total footprint [\tcotwoeq]} \\
\midrule
Train travel within Europe & $2\times \SI{800}{}$ & - & $\SI{5}{nights}$ & $\SI{5}{days}$ & $0.111$\\
Flight travel within Europe & - & $2\times \SI{2.5}{}$ & $\SI{5}{nights}$ & $\SI{5}{days}$ & $0.711$\\
Flight travel across continents & $2\times \SI{275}{}$ & $2\times \SI{12}{}$ & $\SI{14}{nights}$ & $\SI{10}{days}$ & $4.267$\\
\bottomrule
\end{tabular}
\caption{Total footprint in \tcotwoeq obtained for three different scenarios of business trips. The numbers used for the conversion to \tcotwoeq are defined in Table~\ref{BusinessTravelTable}.}
\label{tab:TravScenarios}
\end{table}

If one considers the first scenario, but replaces train with a flight as the mode of transport, which requires \SI{1.5}{hours} for the same distance of \SI{800}{km} one way, the total footprint increases to \SI{0.451}{\tcotwoeq}. The change to flight as transport mode therefore yields a footprint that is $4$ times larger compared to travel by train.

The duration for the last scenario is extended from the typical 5 day duration observed in other scenarios to recognise that longer-distance trips often require more justification to be funded, e.g. longer educational programs. It must be noted however, that the footprint of hotel stay and event venue remains small compared to that of the travel itself. 
This implies that the carbon footprint of a shorter trip would be similar as long as the footprint of the flight travel is not compensated. Note that hotel and venue footprints are based on German numbers and are assumed to be valid in the international context as well. This assumption might be refined in a future version of the Kyf calculator.

\section*{Data availability}
Data used in this study is either contained in the cited reports, or extracted from indicated figures. The extracted numerical numbers are provided in the supplementary material.

\section*{Code availability}
The conducted calculations are fully described in the study. No further code package was used.

\section*{Acknowledgements}
Our warm thanks go to Ruth Jacobs, Srijan Sehgal, and Dima El Khechen from the yHEP management board for their constructive input on this paper and to Michael Lupberger from the yHEP management board for his feedback to the Kyf calculator. Many thanks also go to further current and former yHEP management board members for their support and constructive input to the Kyf campaign efforts: Farah Afzal, Leonel Morejon, Meike Küßner, Felipe Pe\~{n}a, Philipp Krönert, Alina Nasr-Esfahani, Annika Thiel. Additionally, we would like to express our gratitude for feedback on the paper and the calculator to the following colleagues: Lutz Feld for the committee for particle physics (KET) in Germany, Uli Katz for the committee for astro-particle physics (KAT) in Germany, Michael Block for the committee for physics of nucleons and hadrons (KHuK) in Germany, Eric Bründermann for the committee for accelerator physics (KfB) in Germany, Denise Völker as head of the DESY Sustainability group, Sonja Kleiner and Luisa Ulrici as leaders of the CERN Environment group, Frank Schlünzen for computing inputs, and Maria Zeitz from KlimAktiv. We thank KlimAktiv especially for the permission to reference and forward to the UBA carbon calculator for the private footprint.
We also thank the two anonymous reviewers from npj Climate Action for their very valuable comments to our paper.
The authors of this paper take responsibility for any errors in this paper. This study did not receive dedicated funding.

\section*{Author contributions}
The initiative, idea, and scope of the study was created by V.L. All authors planned the structure of the paper. V.L. and N.K.B. developed the experimental footprint, V.L. and P.N. the institute footprint, P.N. and S.G. the computing footprint, and S.G. and N.K.B. the travel footprint. All authors contributed to writing, editing, refinement, and final review of the paper.

\section*{Competing interests}
The authors declare no competing interests.

\bibliography{bibliography}

\begin{thebibliography}{10}
\urlstyle{rm}
\expandafter\ifx\csname url\endcsname\relax
  \def\url#1{\texttt{#1}}\fi
\expandafter\ifx\csname urlprefix\endcsname\relax\def\urlprefix{URL }\fi
\expandafter\ifx\csname doiprefix\endcsname\relax\def\doiprefix{DOI: }\fi
\providecommand{\bibinfo}[2]{#2}
\providecommand{\eprint}[2][]{\url{#2}}

\bibitem{Arrhenius}
\bibinfo{author}{Arrhenius, S.}
\newblock \bibinfo{journal}{\bibinfo{title}{{XXXI. On the influence of carbonic acid in the air upon the temperature of the ground}}}.
\newblock {\emph{\JournalTitle{The London, Edinburgh, and Dublin Philosophical Magazine and Journal of Science}}} \textbf{\bibinfo{volume}{41}}, \bibinfo{pages}{237--276}, \doiprefix\url{10.1080/14786449608620846} (\bibinfo{year}{1896}).

\bibitem{KeelingCurve}
\bibinfo{author}{{UC San Diego and SCRIPPS Institution of Oceanography}}.
\newblock \bibinfo{title}{{The Keeling Curve - Full Record}}.
\newblock \bibinfo{howpublished}{\url{https://keelingcurve.ucsd.edu}} (\bibinfo{year}{2023}).
\newblock \bibinfo{note}{Accessed: 2023-10-22}.

\bibitem{co2inlast800kyears}
\bibinfo{author}{Lüthi, D.} \emph{et~al.}
\newblock \bibinfo{journal}{\bibinfo{title}{{High-resolution carbon dioxide concentration record 650,000–800,000 years before present}}}.
\newblock {\emph{\JournalTitle{Nature}}} \textbf{\bibinfo{volume}{453}}, \bibinfo{pages}{379--382}, \doiprefix\url{https://doi.org/10.1038/nature06949} (\bibinfo{year}{2008}).

\bibitem{CopernicusTemperatureData}
\bibinfo{author}{{Copernicus Climate Change Service/ECMWF}}.
\newblock \bibinfo{title}{{Surface air temperature for December 2023}}.
\newblock \bibinfo{howpublished}{\url{https://climate.copernicus.eu/surface-air-temperature-december-2023}} (\bibinfo{year}{2023}).
\newblock \bibinfo{note}{Accessed: 2024-02-09}.

\bibitem{IPPCAR6SPM}
\bibinfo{author}{{Intergovernmental Panel on Climate Change (IPCC)}}.
\newblock \emph{\bibinfo{title}{{Climate Change 2021 – The Physical Science Basis: Working Group I Contribution to the Sixth Assessment Report of the Intergovernmental Panel on Climate Change}}}, chap. \bibinfo{chapter}{{Summary for Policymakers}}, \bibinfo{pages}{3--32} (\bibinfo{publisher}{Cambridge University Press}, \bibinfo{year}{2023}).

\bibitem{ParisAgreement}
\bibinfo{author}{{United Nations}}.
\newblock \bibinfo{journal}{\bibinfo{title}{{Paris Agreement}}}.
\newblock {\emph{\JournalTitle{United Nations Treaty Series}}} \textbf{\bibinfo{volume}{3156}}, \bibinfo{pages}{79} (\bibinfo{year}{2015}).

\bibitem{UNpopulationData}
\bibinfo{author}{{United Nations}}.
\newblock \bibinfo{title}{{Global issues: Population}}.
\newblock \bibinfo{howpublished}{\url{https://www.un.org/en/global-issues/population}} (\bibinfo{year}{2023}).
\newblock \bibinfo{note}{Accessed: 2023-10-22}.

\bibitem{UBAemissionsGermany}
\bibinfo{author}{Schächtele, K.} \& \bibinfo{author}{Hertle, H.}
\newblock \bibinfo{journal}{\bibinfo{title}{{Die \cotwo Bilanz des Bürgers - Recherche für ein internetbasiertes Tool zur Erstellung persönlicher CO2 Bilanzen}}}.
\newblock {\emph{\JournalTitle{Umweltbundesamt}}} \textbf{\bibinfo{volume}{Förderkennzeichen: 206 42 110}} (\bibinfo{year}{2007}).

\bibitem{UBAcalculatorRef}
\bibinfo{author}{Schunkert, S.} \emph{et~al.}
\newblock \bibinfo{journal}{\bibinfo{title}{{Der UBA-\cotwo-Rechner für Privatpersonen - Hintergrundinformationen}}}.
\newblock {\emph{\JournalTitle{Umweltbundesamt}}} \textbf{\bibinfo{volume}{Forschungskennzahl: 3718 16 313 0}} (\bibinfo{year}{2022}).

\bibitem{CERN1718}
\bibinfo{author}{{CERN}}.
\newblock \bibinfo{title}{{CERN Environment Report 2017–2018}}, \doiprefix\url{10.25325/CERN-Environment-2020-001} (\bibinfo{year}{2020}).

\bibitem{CERN1920}
\bibinfo{author}{{CERN}}.
\newblock \bibinfo{title}{{CERN Environment Report 2019–2020}}, \doiprefix\url{10.25325/CERN-Environment-2021-002} (\bibinfo{year}{2021}).

\bibitem{CERN2122}
\bibinfo{author}{{CERN}}.
\newblock \bibinfo{title}{{CERN Environment Report 2021–2022}}, \doiprefix\url{10.25325/CERN-Environment-2023-003} (\bibinfo{year}{2023}).

\bibitem{DESY1921}
\bibinfo{author}{{DESY}}.
\newblock \bibinfo{title}{{First Sustainability report}}.
\newblock \bibinfo{howpublished}{\url{https://nachhaltigkeit.desy.de/sustainability_report/index_eng.html}} (\bibinfo{year}{2022}).

\bibitem{LCA4ILCandCLIC}
\bibinfo{author}{Evans, S.} \emph{et~al.}
\newblock \bibinfo{title}{{Life Cycle Assessment - Comparative environmental footprint for future linear colliders CLIC and ILC}}.
\newblock \bibinfo{howpublished}{\url{https://edms.cern.ch/ui/file/2917948/1/Life_Cycle_Assessment_for_CLIC_and_ILC_Final_Report_July_2023.pdf}} (\bibinfo{year}{2023}).

\bibitem{FCCfootprint}
\bibinfo{author}{Mauree, D.}
\newblock \bibinfo{title}{{FCC – Carbon budget study}}.
\newblock \bibinfo{howpublished}{\url{https://indico.cern.ch/event/1298458/contributions/5981426/attachments/2876462/5037578/100196.09_FCC_Carbon_budget.pdf}} (\bibinfo{year}{2024}).
\newblock \bibinfo{note}{FCC Week}.

\bibitem{CEPCfootprint}
\bibinfo{author}{Li, Y.}
\newblock \bibinfo{title}{{Efforts to make CEPC a green machine}}.
\newblock \bibinfo{howpublished}{\url{https://indico.cern.ch/event/1355767/contributions/5967009/attachments/2874096/5034223/Efforts\%20to\%20make\%20CEPC\%20a\%20green\%20machine.pdf}} (\bibinfo{year}{2024}).
\newblock \bibinfo{note}{Sustainable HEP2024 - 3rd edition}.

\bibitem{C3footprint}
\bibinfo{author}{Breidenbach, M.}, \bibinfo{author}{Bullard, B.}, \bibinfo{author}{Nanni, E.~A.}, \bibinfo{author}{Ntounis, D.} \& \bibinfo{author}{Vernieri, C.}
\newblock \bibinfo{journal}{\bibinfo{title}{{Sustainability Strategy for the Cool Copper Collider}}}.
\newblock {\emph{\JournalTitle{PRX Energy}}} \textbf{\bibinfo{volume}{2}}, \bibinfo{pages}{047001}, \doiprefix\url{10.1103/PRXEnergy.2.047001} (\bibinfo{year}{2023}).

\bibitem{ALLEA}
\bibinfo{author}{{ALLEA}}.
\newblock \bibinfo{title}{{Towards Climate Sustainability of the Academic System in Europe and Beyond}}.
\newblock \bibinfo{howpublished}{\url{https://allea.org/wp-content/uploads/2022/05/ALLEA-Report-Towards-Climate-Sustainability-of-the-Academic-System.pdf}} (\bibinfo{year}{2022}).

\bibitem{HECAPplus}
\bibinfo{author}{Banerjee, S.} \emph{et~al.}
\newblock \bibinfo{title}{{Environmental sustainability in basic research: a perspective from HECAP+}} (\bibinfo{year}{2023}).
\newblock \eprint{arXiv:2306.02837}.

\bibitem{SnowmassSustInput}
\bibinfo{author}{Bloom, K.} \emph{et~al.}
\newblock \bibinfo{title}{{Climate impacts of particle physics}}.
\newblock In \emph{\bibinfo{booktitle}{{Snowmass 2021}}} (\bibinfo{year}{2022}).
\newblock \eprint{arXiv:2203.12389}.

\bibitem{SustFutureAcc}
\bibinfo{author}{Bloom, K.} \& \bibinfo{author}{Boisvert, V.}
\newblock \bibinfo{title}{{Sustainability and Carbon Emissions of Future Accelerators}}, \doiprefix\url{10.1146/annurev-nucl-121423-100906} (\bibinfo{year}{2024}).
\newblock \eprint{arXiv:2411.03473}.

\bibitem{AstroInLowCarbonFuture}
\bibinfo{author}{Matzner, C.~D.} \emph{et~al.}
\newblock \bibinfo{title}{{Astronomy in a Low-Carbon Future}}, \doiprefix\url{10.5281/zenodo.3758549} (\bibinfo{year}{2019}).
\newblock \eprint{arXiv:1910.01272}.

\bibitem{Labos1point5}
\bibinfo{author}{{Labos 1point5}}.
\newblock \bibinfo{title}{{Reducing the environmental footprint of our research activities}}.
\newblock \bibinfo{howpublished}{\url{https://labos1point5.org}}.
\newblock \bibinfo{note}{Accessed: December 2024}.

\bibitem{GES1point5}
\bibinfo{author}{Mariette, J.} \emph{et~al.}
\newblock \bibinfo{journal}{\bibinfo{title}{An open-source tool to assess the carbon footprint of research}}.
\newblock {\emph{\JournalTitle{Environmental Research: Infrastructure and Sustainability}}} \textbf{\bibinfo{volume}{2}}, \bibinfo{pages}{035008}, \doiprefix\url{10.1088/2634-4505/ac84a4} (\bibinfo{year}{2022}).

\bibitem{AstrMaxPlanckInstEmissions}
\bibinfo{author}{Jahnke, K.} \emph{et~al.}
\newblock \bibinfo{journal}{\bibinfo{title}{{An astronomical institute’s perspective on meeting the challenges of the climate crisis}}}.
\newblock {\emph{\JournalTitle{Nat Astron}}} \textbf{\bibinfo{volume}{4}}, \bibinfo{pages}{812--815}, \doiprefix\url{10.1038/s41550-020-1202-4} (\bibinfo{year}{2020}).

\bibitem{AstrFutureCarbonEmissions}
\bibinfo{author}{Knödlseder, J.}, \bibinfo{author}{Coriat, M.}, \bibinfo{author}{Garnier, P.} \& \bibinfo{author}{Hughes, A.}
\newblock \bibinfo{journal}{\bibinfo{title}{Scenarios of future annual carbon footprints of astronomical research infrastructures}}.
\newblock {\emph{\JournalTitle{Nature Astronomy}}} \textbf{\bibinfo{volume}{8}}, \bibinfo{pages}{1478–1486}, \doiprefix\url{10.1038/s41550-024-02346-0} (\bibinfo{year}{2024}).

\bibitem{yHEPwebpage}
\bibinfo{author}{{young High Energy Physicists Association (yHEP)}}.
\newblock \bibinfo{title}{{Know your footprint}}.
\newblock \bibinfo{howpublished}{\url{https://yhep.desy.de/sustainability/}}.
\newblock \bibinfo{note}{Accessed: December 2023}.

\bibitem{UBAcalculator}
\bibinfo{author}{{Umweltbundesamt}}.
\newblock \bibinfo{title}{{Carbon Calculator}}.
\newblock \bibinfo{howpublished}{\url{https://uba.co2-rechner.de/en_GB/}} (\bibinfo{year}{2023}).
\newblock \bibinfo{note}{Accessed: 2023-11-01}.

\bibitem{KlimAktiv}
\bibinfo{author}{{KlimAktiv}}.
\newblock \bibinfo{title}{{Unsere Mission: Null Emission}}.
\newblock \bibinfo{howpublished}{\url{https://klimaktiv.de}} (\bibinfo{year}{2024}).
\newblock \bibinfo{note}{Accessed: 2024-09-11}.

\bibitem{LHCAcc}
\bibinfo{author}{Evans, L.} \& \bibinfo{author}{Bryant, P.}
\newblock \bibinfo{journal}{\bibinfo{title}{{LHC Machine}}}.
\newblock {\emph{\JournalTitle{JINST}}} \textbf{\bibinfo{volume}{3}}, \bibinfo{pages}{S08001}, \doiprefix\url{10.1088/1748-0221/3/08/S08001} (\bibinfo{year}{2008}).

\bibitem{CERNHome}
\bibinfo{author}{CERN}.
\newblock \bibinfo{title}{{Home | CERN}}.
\newblock \bibinfo{howpublished}{\url{https://home.cern/}}.
\newblock \bibinfo{note}{Accessed: 2024-02-09}.

\bibitem{ATLASDet}
\bibinfo{author}{{ATLAS Collaboration}}.
\newblock \bibinfo{journal}{\bibinfo{title}{{The ATLAS Experiment at the CERN Large Hadron Collider}}}.
\newblock {\emph{\JournalTitle{JINST}}} \textbf{\bibinfo{volume}{3}}, \bibinfo{pages}{S08003}, \doiprefix\url{10.1088/1748-0221/3/08/S08003} (\bibinfo{year}{2008}).

\bibitem{CMSDet}
\bibinfo{author}{{CMS Collaboration}}.
\newblock \bibinfo{journal}{\bibinfo{title}{{The CMS Experiment at the CERN LHC}}}.
\newblock {\emph{\JournalTitle{JINST}}} \textbf{\bibinfo{volume}{3}}, \bibinfo{pages}{S08004}, \doiprefix\url{10.1088/1748-0221/3/08/S08004} (\bibinfo{year}{2008}).

\bibitem{ALICEDet}
\bibinfo{author}{{ALICE Collaboration}}.
\newblock \bibinfo{journal}{\bibinfo{title}{{The ALICE experiment at the CERN LHC}}}.
\newblock {\emph{\JournalTitle{JINST}}} \textbf{\bibinfo{volume}{3}}, \bibinfo{pages}{S08002}, \doiprefix\url{10.1088/1748-0221/3/08/S08002} (\bibinfo{year}{2008}).

\bibitem{LHCbDet}
\bibinfo{author}{{LHCb Collaboration}}.
\newblock \bibinfo{journal}{\bibinfo{title}{{The LHCb Detector at the LHC}}}.
\newblock {\emph{\JournalTitle{JINST}}} \textbf{\bibinfo{volume}{3}}, \bibinfo{pages}{S08005}, \doiprefix\url{10.1088/1748-0221/3/08/S08005} (\bibinfo{year}{2008}).

\bibitem{LHCbTDR}
\bibinfo{author}{{LHCb Collaboration}}.
\newblock \bibinfo{title}{{Framework TDR for the LHCb Upgrade II: Opportunities in flavour physics, and beyond, in the HL-LHC era}}.
\newblock \bibinfo{howpublished}{{CERN-LHCC-2021-012, LHCB-TDR-023}} (\bibinfo{year}{2021}).

\bibitem{Plotdigitzer}
\bibinfo{title}{{PlotDigitizer: Version 3.1.5}}.
\newblock \bibinfo{howpublished}{\url{https://plotdigitizer.com}} (\bibinfo{year}{2024}).

\bibitem{ALICEmemb}
\bibinfo{author}{{ALICE Collaboration}}.
\newblock \bibinfo{title}{{Collaboration Map}}.
\newblock \bibinfo{howpublished}{\url{https://alice-collaboration.web.cern.ch/collaboration/map}} (\bibinfo{year}{2024}).
\newblock \bibinfo{note}{Accessed: 2024-01-05}.

\bibitem{LHCbmemb}
\bibinfo{author}{{LHCb Collaboration}}.
\newblock \bibinfo{title}{{The LHCb Collaboration}}.
\newblock \bibinfo{howpublished}{\url{https://lhcb-outreach.web.cern.ch/collaboration/}} (\bibinfo{year}{2021}).
\newblock \bibinfo{note}{Accessed: 2024-01-05}.

\bibitem{CMSmemb}
\bibinfo{author}{{CMS Collaboration}}.
\newblock \bibinfo{title}{{People Statistics}}.
\newblock \bibinfo{howpublished}{\url{https://cms.cern/index.php/collaboration/people-statistics}} (\bibinfo{year}{2022}).
\newblock \bibinfo{note}{Accessed: 2024-01-05}.

\bibitem{ATLASmemb}
\bibinfo{author}{{ATLAS Collaboration}}.
\newblock \bibinfo{title}{{The Collaboration}}.
\newblock \bibinfo{howpublished}{\url{https://atlas.cern/Discover/Collaboration}} (\bibinfo{year}{2024}).
\newblock \bibinfo{note}{Accessed: 2024-01-05}.

\bibitem{DESYenergy}
\bibinfo{author}{{DESY}}.
\newblock \bibinfo{title}{{Energy monitoring}}.
\newblock \bibinfo{howpublished}{\url{https://nachhaltigkeit.desy.de/energy_management/energy_monitoring/index_eng.html}} (\bibinfo{year}{2023}).
\newblock \bibinfo{note}{Accessed: 2024-01-07}.

\bibitem{ElectricityMaps}
\bibinfo{author}{{Electricity maps}}.
\newblock \bibinfo{title}{{Germany}}.
\newblock \bibinfo{howpublished}{\url{https://app.electricitymaps.com/zone/DE}} (\bibinfo{year}{2023}).
\newblock \bibinfo{note}{Accessed: 2024-01-07}.

\bibitem{ESO2020}
\bibinfo{author}{{European Organisation for Astronomical Research in the Southern Hemisphere}}.
\newblock \bibinfo{title}{{ESO Annual Report 2020}}.
\newblock \bibinfo{howpublished}{\url{https://www.eso.org/public/products/annualreports/ar_2020/}} (\bibinfo{year}{2021}).
\newblock \bibinfo{note}{Accessed: 2023-10-16}.

\bibitem{ESO2021}
\bibinfo{author}{{European Organisation for Astronomical Research in the Southern Hemisphere}}.
\newblock \bibinfo{title}{{ESO Annual Report 2021}}.
\newblock \bibinfo{howpublished}{\url{https://www.eso.org/public/products/annualreports/ar_2021/}} (\bibinfo{year}{2022}).
\newblock \bibinfo{note}{Accessed: 2023-10-16}.

\bibitem{ESO2022}
\bibinfo{author}{{European Organisation for Astronomical Research in the Southern Hemisphere}}.
\newblock \bibinfo{title}{{ESO Annual Report 2022}}.
\newblock \bibinfo{howpublished}{\url{https://www.eso.org/public/products/annualreports/ar_2022/}} (\bibinfo{year}{2023}).
\newblock \bibinfo{note}{Accessed: 2023-10-23}.

\bibitem{ESOpubpage}
\bibinfo{author}{{European Southern Observatory}}.
\newblock \bibinfo{title}{{About ESO}}.
\newblock \bibinfo{howpublished}{\url{https://www.eso.org/public/about-eso/}} (\bibinfo{year}{2023}).
\newblock \bibinfo{note}{Accessed: 2024-01-03}.

\bibitem{UniWeimar}
\bibinfo{author}{{Bauhaus-University Weimar}}.
\newblock \bibinfo{title}{{Environmental report 2023}}.
\newblock \bibinfo{howpublished}{\url{https://www.uni-weimar.de/en/university/profile/sustainable-university/environmental-reports/}} (\bibinfo{year}{2023}).

\bibitem{UniTuebingen}
\bibinfo{author}{{Eberhard Karls Universität Tübingen}}.
\newblock \bibinfo{title}{{Umwelterklärung 2023}}.
\newblock \bibinfo{howpublished}{\url{https://uni-tuebingen.de/einrichtungen/verwaltung/viii-bau-arbeitssicherheit-und-umwelt/abteilung-3-umwelt-energie-und-klima/downloads/}} (\bibinfo{year}{2023}).

\bibitem{UniHann}
\bibinfo{author}{{Leibniz University Hannover}}.
\newblock \bibinfo{title}{{Umweltbericht 2017-2019}}.
\newblock \bibinfo{howpublished}{\url{https://www.uni-hannover.de/fileadmin/luh/content/webredaktion/universitaet/publikationen/umweltbericht/umweltbericht_17_19.pdf}} (\bibinfo{year}{2021}).

\bibitem{UniFR}
\bibinfo{author}{{University of Freiburg}}.
\newblock \bibinfo{title}{{Umweltbericht 2019/2020}}.
\newblock \bibinfo{howpublished}{\url{https://www.nachhaltige.uni-freiburg.de/de/umweltberichte/umweltbericht-2019-2020-2}} (\bibinfo{year}{2021}).

\bibitem{CERNpop}
\bibinfo{author}{CERN}.
\newblock \bibinfo{title}{{CERN Annual Personnel Statistics 2022}}.
\newblock \bibinfo{howpublished}{\url{https://cds.cern.ch/record/2858688}} (\bibinfo{year}{2022}).

\bibitem{bruers2023resourceaware}
\bibinfo{author}{Bruers, B.} \emph{et~al.}
\newblock \bibinfo{journal}{\bibinfo{title}{{Resource-aware Research on Universe and Matter: Call-to-Action in Digital Transformation}}}.
\newblock {\emph{\JournalTitle{Eur. Phys. J. Spec. Top}}}  (\bibinfo{year}{2024}).
\newblock \bibinfo{note}{Accepted 29 November 2024}, \eprint{arXiv:2311.01169}.

\bibitem{PUE}
\bibinfo{author}{{The green grid}}.
\newblock \bibinfo{title}{{PUE: A Comprehensive Examination of the Metric}}.
\newblock \bibinfo{howpublished}{\url{https://www.thegreengrid.org/en/resources/library-and-tools/20-PUE:-A-Comprehensive-Examination-of-the-Metric}} (\bibinfo{year}{2014}).
\newblock \bibinfo{note}{Accessed: 2024-01-16}.

\bibitem{cuda}
\bibinfo{author}{{Tech Centurion}}.
\newblock \bibinfo{title}{{Nvidia CUDA Cores Explained}}.
\newblock \bibinfo{howpublished}{\url{https://www.techcenturion.com/nvidia-cuda-cores/}}.
\newblock \bibinfo{note}{Accessed: 2024-03-04}.

\bibitem{EnEfG}
\bibinfo{author}{{Bundesministerium der Justiz}}.
\newblock \bibinfo{title}{{Gesetz zur Steigerung der Energieeffizienz in Deutschland}}.
\newblock \bibinfo{howpublished}{\url{https://www.gesetze-im-internet.de/enefg/BJNR1350B0023.html}}.
\newblock \bibinfo{note}{Effective date: 18 November 2023}.

\bibitem{Hawk}
\bibinfo{author}{HLRS}.
\newblock \bibinfo{title}{{HPE Apollo (Hawk)}}.
\newblock \bibinfo{howpublished}{\url{https://www.hlrs.de/solutions/systems/hpe-apollo-hawk}}.
\newblock \bibinfo{note}{Accessed: 2024-01-24}.

\bibitem{DESYMaxwell_private_communication}
\bibinfo{author}{Schluenzen, F.}
\newblock \bibinfo{title}{{Computing Cluster Power Consumption Data}} (\bibinfo{year}{2023}).
\newblock \bibinfo{note}{Private communication}.

\bibitem{NVIDIA_A100_GPU}
\bibinfo{author}{Wifocusonenergy}.
\newblock \bibinfo{title}{{How much electricity does the A100 GPU use?}}
\newblock \bibinfo{howpublished}{\url{https://wifocusonenergy.com/optimizing-power-consumption-understanding-the-nvidia-a100s-energy-usage}}.
\newblock \bibinfo{note}{Accessed: 2024-01-16}.

\bibitem{Big_Panda}
\bibinfo{author}{BigPanDA}.
\newblock \bibinfo{title}{{PanDA dashboard}}.
\newblock \bibinfo{howpublished}{\url{https://bigpanda.cern.ch/dash/region/}}.
\newblock \bibinfo{note}{Accessed: 2024-01-16}.

\bibitem{Panda_API}
\bibinfo{author}{BigPanDA}.
\newblock \bibinfo{title}{{PanDA system python API reference}}.
\newblock \bibinfo{howpublished}{\url{https://panda-wms.readthedocs.io/en/latest/client/rest.html}} (\bibinfo{year}{2020}).
\newblock \bibinfo{note}{Accessed: 2024-01-16}.

\bibitem{onedrive}
\bibinfo{author}{Microsoft}.
\newblock \bibinfo{title}{{Microsoft OneDrive}}.
\newblock \bibinfo{howpublished}{\url{https://www.microsoft.com/en-us/microsoft-365/onedrive/online-cloud-storage}}.

\bibitem{googledrive}
\bibinfo{author}{Google}.
\newblock \bibinfo{title}{{Google Drive}}.
\newblock \bibinfo{howpublished}{\url{https://www.google.com/drive/}}.

\bibitem{onedrive_dashboard}
\bibinfo{author}{Microsoft}.
\newblock \bibinfo{title}{{Microsoft OneDrive Emisions Impact Dashboard}}.
\newblock \bibinfo{howpublished}{\url{https://www.microsoft.com/en-us/sustainability/emissions-impact-dashboard}}.

\bibitem{googledrive_footprintcalculator}
\bibinfo{author}{Google}.
\newblock \bibinfo{title}{{Google Drive Footprint Calculator}}.
\newblock \bibinfo{howpublished}{\url{https://cloud.google.com/carbon-footprint}}.

\bibitem{Umweltbundesamt_TREMOD}
\bibinfo{author}{{Umweltbundesamt}}.
\newblock \bibinfo{title}{{Emissionsdaten}}.
\newblock \bibinfo{howpublished}{\url{https://www.umweltbundesamt.de/themen/verkehr/emissionsdaten}} (\bibinfo{year}{2023}).
\newblock \bibinfo{note}{Accessed: 2024-02-12}.

\bibitem{VeranstaltungRechnerUBA}
\bibinfo{author}{Umweltbundesamt}.
\newblock \bibinfo{title}{{UBA CO2-Rechner für Veranstaltungen (2024)}}.
\newblock \bibinfo{howpublished}{\url{https://uba-event-free.co2-rechner.pro/de_DE/}} (\bibinfo{year}{2024}).
\newblock \bibinfo{note}{Accessed: 2024-03-02}.

\bibitem{EnergyProvHann}
\bibinfo{author}{{Technische Universität Dresden}}.
\newblock \bibinfo{title}{{Certification for the district heating system from enercity AG Hannover}}.
\newblock \bibinfo{howpublished}{\url{https://www.enercity.de/assets/cms/enercity-de/Privatkunden/Produkte/Waerme/Fernwaerme/Downloads/FW309-6.pdf}} (\bibinfo{year}{2022}).

\bibitem{WaterFootprint}
\bibinfo{author}{{GUTcert}}.
\newblock \bibinfo{title}{{Vergleich des \cotwo-Fußabdrucks von Mineral- und Trinkwasser}}.
\newblock \bibinfo{howpublished}{\url{https://atiptap.org/files/studie_gutcert_pcf_wasser.pdf}} (\bibinfo{year}{2020}).

\bibitem{ProcurementMasterThesis}
\bibinfo{author}{Eichler, M.}
\newblock \bibinfo{title}{{Evaluating Environmental Impacts of University Procurements – An Environmentally Extended Multiregional Input Output Analysis of Albert-Ludwigs-Universität Freiburg}} (\bibinfo{year}{2020}).

\bibitem{ProcurementUniFR}
\bibinfo{author}{Pauliuk, S.} \emph{et~al.}
\newblock \bibinfo{journal}{\bibinfo{title}{{Treibhausgasbilanz der Universität Freiburg im Breisgau 2017}}}.
\newblock {\emph{\JournalTitle{Industrial Ecology Freiburg (IEF) Working Paper}}} \textbf{\bibinfo{volume}{1}}, \doiprefix\url{10.6094/UNIFR/176419} (\bibinfo{year}{2021}).

\end{thebibliography}

\section*{Additional information}

\textbf{Supplementary Information} The online version contains supplementary material available at the published link and below. \newline
\textbf{Correspondence and requests} for materials should be addressed to V.L. \newline
\textbf{Reprints and permissions information} is available at \href{www.nature.com/reprints}{www.nature.com/reprints}. \newline
\textbf{Publisher’s note} Springer Nature remains neutral with regard to jurisdictional claims in published maps and institutional affiliations.
\begin{wrapfigure}[2]{l}{0.08\textwidth}
\centering
    \vspace{-0.4cm}
    \includegraphics[width=0.10\textwidth]{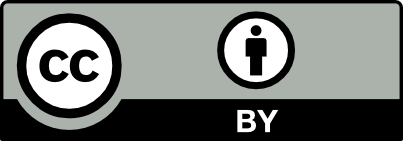}
\end{wrapfigure}
\textbf{Open Access} This article is licensed under a Creative Commons Attribution 4.0 International License, which permits use, sharing, adaptation, distribution and reproduction in any medium or format, as long as you give appropriate credit to the original author(s) and the source, provide a link to the
Creative Commons licence, and indicate if changes were made. The images or other third party material in this article are included in the article’s Creative Commons licence, unless indicated otherwise in a credit line to the
material. If material is not included in the article’s Creative Commons licence and your intended use is not permitted by statutory regulation or exceeds the permitted use, you will need to obtain permission directly from
the copyright holder. To view a copy of this licence, visit \href{http://creativecommons.org/licenses/by/4.0/}{http://creativecommons.org/licenses/by/4.0/}.

\clearpage

\appendix
\section*{\LARGE{Supplementary material}}

\section{Digitisation of CERN scope 1 emissions}
\label{app:CERNscope1SplitUp}

\begin{figure}[hbt]
    \centering
    \includegraphics[width=0.40\textwidth]{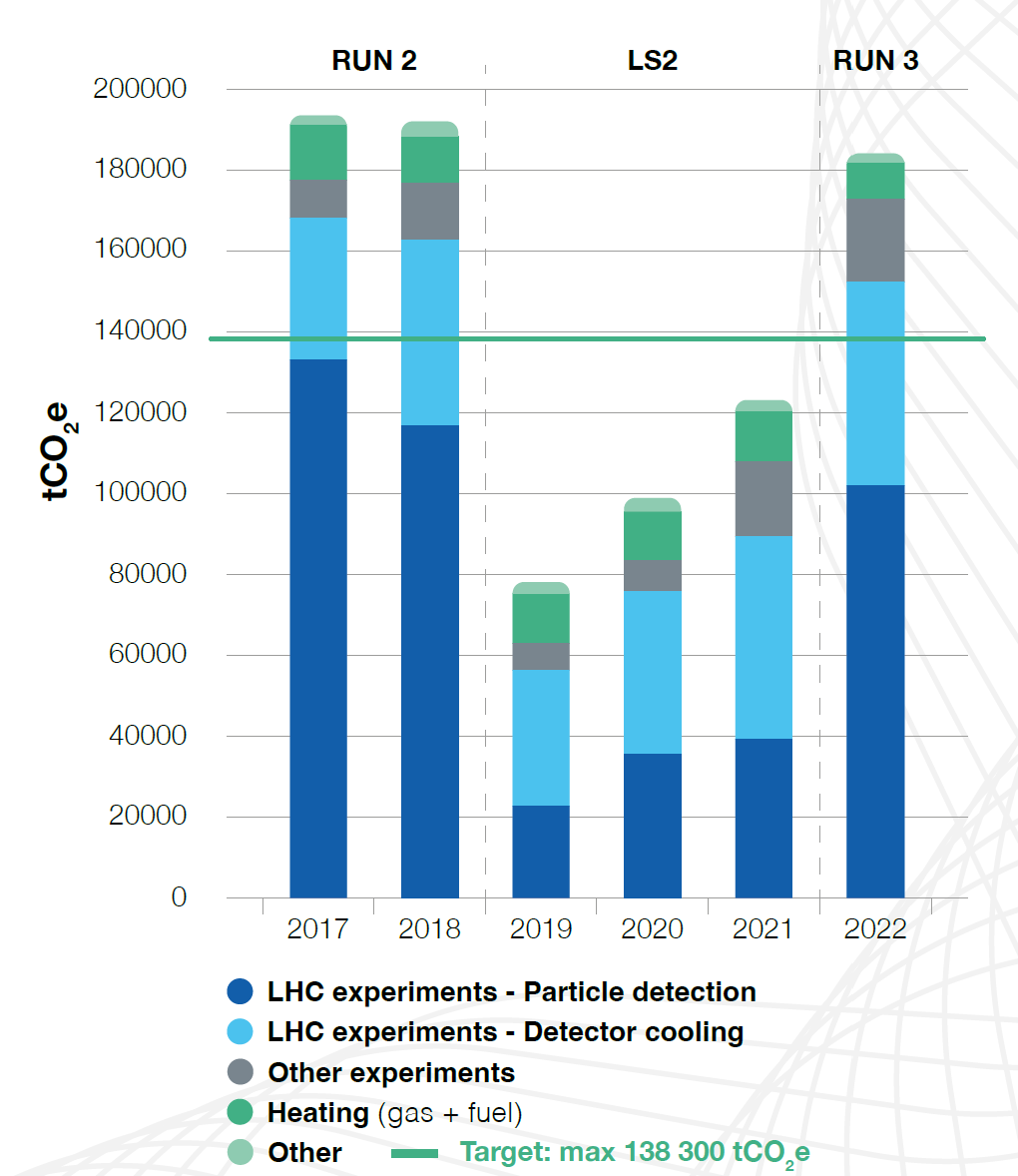}
    \caption{CERN scope 1 emissions for 2017--2022 by category~\cite{CERN2122}. The category \emph{Other} includes air conditioning, electrical insulation, emergency generators and the fuel consumption of the CERN vehicle fleet.}
    \label{fig:CERNscope1}
\end{figure}

Numbers from the digitisation of the figure: \emph{CERN scope~1 emissions for 2017--2022 by category}, from page 18 of the CERN environmental report 2021--2022~\cite{CERN2122}, reproduced in Figure~\ref{fig:CERNscope1}, can be found in Table~\ref{tab:CERNscope1}. The total scope 1 emissions from 2022 determined with \texttt{PlotDigitizer}~\cite{Plotdigitzer} agree within \SI{0.05}{\%} with the numerical value for the total 2022 scope~1 emissions stated in the environmental report. Given resolution precision, this is considered sufficiently close.

\begin{table*}[hbt]
\centering
\begin{tabular}{l r}
\toprule
\textbf{Category}   & \multicolumn{1}{c}{\textbf{Cumulative emissions [\tcotwoeq]}}    \\
\midrule
LHC experiments: Particle detection & \SI{101981}{}   \\
LHC experiments: Detector cooling   & \SI{152444}{}    \\
Other experiments       & \SI{172923}{}    \\
Heating (gas+fuel)      & \SI{181857}{}      \\
Other                   & \SI{184090}{}     \\
\midrule
Total (specified in the report) & \SI{184173}{} \\
\bottomrule
\end{tabular}
\caption{Digitised values for 2022 of the CERN scope 1 emissions, extracted with \texttt{PlotDigitizer}~\cite{Plotdigitzer} from Figure~\ref{fig:CERNscope1}. The values correspond to the upper edge of the bar including the specified category. The value for \emph{Other} should therefore correspond to the total scope 1 emissions value specified in the CERN environmental report 2021--2022~\cite{CERN2122}, listed under \emph{Total (specified in the report)}, which - given resolution precision - is sufficiently close.}
\label{tab:CERNscope1}
\end{table*}

\section{DESY energy consumption}
\label{app:DESYenergy}

\begin{figure*}[h]
    \centering
    \includegraphics[width=0.7\textwidth]{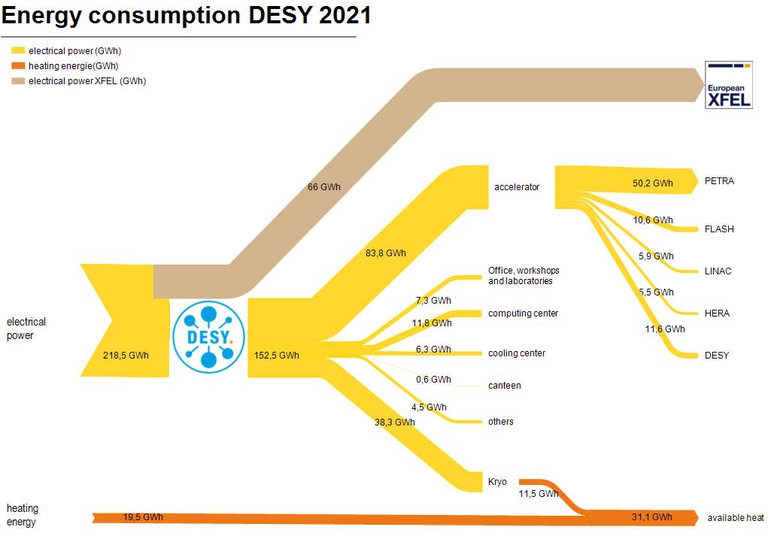}
    \caption{DESY energy consumption in 2021~\cite{DESYenergy}. Electrical power consumption (yellow) is split into consumption by the European XFEL (brown) and DESY (yellow). Heat consumption (orange) contains a contribution from heat recovery from cryogenics plants which is provided for heating and hot water supply to buildings on the DESY campus.}
    \label{fig:DESYenergy}
\end{figure*}

The DESY webpage provides a figure on the energy consumption at DESY in the year 2021~\cite{DESYenergy} which is reproduced in Figure~\ref{fig:DESYenergy}, for permanent availability together with this study.

\section{Leibniz University Hannover}
\label{UniHannover}

The environmental footprint of the Leibniz University Hannover is estimated based on the environmental report for 2017--2019~\cite{UniHann}, using numbers from 2019 for direct comparison to the numbers from the University of Freiburg in the main paper. Numbers for the Leibniz University Hannover, provided in the environmental report, largely specify the original consumption, hence the conversion to \tcotwoeq is performed in this study.

Emissions for the Leibniz University Hannover are provided for the categories: Electricity, Heating/Cooling, Water, and Waste. Numbers for procurement are not provided. While the total footprint is therefore not comparable to the one from the University of Freiburg, for the categories except procurement, the numbers from the Leibniz University Hannover provide an interesting cross-check.

Consumption numbers, as provided by the Leibniz University Hannover, conversion factors and \cotwo emissions are listed in Table~\ref{tab:UniHannSplit}. The conversion factors for green (conventional) electricity are used as \SI{35}{g\cotwo e/kWh} (\SI{416}{g\cotwo e/kWh})~\cite{ElectricityMaps}. The conversion factor of \SI{98.7}{kg\cotwo e/MWh}~\cite{EnergyProvHann} (including green house gas emissions other than \cotwo) for heating is obtained from the energy provider \emph{enercity} of the Leibniz University Hannover, which is specified to use district heating. The climate impact for water consumption assumes that the used water is drinking water, and the conversion factor of \SI{0.35}{g\cotwo e/l} is obtained from a study comparing the climate impact of bottled vs. tap water~\cite{WaterFootprint}. The total amount of waste in 2019, as well as the amount of toxic waste (\emph{Sonderabfall}) in 2019, are read off from two figures contained in the environmental report 2017--2019 by the Leibniz University Hannover~\cite{UniHann}. The conversion factors for the \cotwo emissions for waste are not provided in that report, however, in the environmental report by the University of Freiburg~\cite{UniFR} sufficient information is provided to calculate conversion factors: For the University of Freiburg, \SI{88}{t} of toxic waste are specified to correspond to \SI{24}{\%} of the total amount of emissions from waste, which are given as \SI{577}{\tcotwoeq}. This results in a conversion factor of \SI{1.57}{\tcotwoeq/t} for toxic waste. The remaining \SI{1019}{t} of waste for the University of Freiburg are specified to generate \SI{76}{\%} of the emissions, resulting in a conversion factor of \SI{0.43}{\tcotwoeq/t} for other waste. These conversion factors are transferred to the waste of the Leibniz University Hannover, assuming implicitly a similar waste composition within the two categories and a similar treatment by the two universities.

\begin{table}[tb]
\centering
\begin{tabular}{l r r r}
\toprule
\textbf{Category}   & \multicolumn{1}{c}{\textbf{Consumption}} & \multicolumn{1}{c}{\textbf{Conversion factor}} & \multicolumn{1}{c}{\textbf{Emissions [\tcotwoeq]}}    \\
\midrule
Electricity     & \SI{58.66}{GWh}  & \SI{35}{g\cotwo e/kWh} (\SI{416}{g\cotwo e/kWh})~\cite{ElectricityMaps} & \SI{2053}{} (\SI{24403}{})  \\
Heating/Cooling & \SI{50.01}{GWh} & \SI{98.7}{kg\cotwo e/MWh}~\cite{EnergyProvHann} & \SI{4936}{}  \\
Water           & \SI{696000}{m^3}  & \SI{0.35}{g/l}~\cite{WaterFootprint}  & \SI{244}{}  \\
Waste           & \SI{1620}{t}     \\
... Toxic waste & \SI{63}{t}  & \SI{1.57}{\tcotwoeq/t}~\cite{UniFR}  & \SI{99}{}    \\
... Other waste & \SI{1557}{t} & \SI{0.43}{\tcotwoeq/t}~\cite{UniFR} & \SI{670}{}     \\
\midrule
Total           & & & \SI{8001}{} (\SI{30351}{}) \\
\bottomrule
\end{tabular}
\caption{Breakdown of the total \cotwo emissions of the Leibniz University Hannover. Consumption values are taken from the environmental report 2017--2019~\cite{UniHann}. Electricity values are provided nominally using green electricity supply. The values in brackets provide the numbers using electricity produced with the German electricity mix. Sources for the conversion factors are indicated, and further explained in the text. Note that the sums are calculated with more decimal places than indicated, resulting in small differences to the sums of the listed values.}
\label{tab:UniHannSplit}
\end{table}

The total emissions for the Leibniz University Hannover are estimated to be \SI{8001}{\tcotwoeq} (\SI{30351}{\tcotwoeq}), with green (conventional) electricity production. Comparing only the equivalent categories, the emissions of \SI{8001}{\tcotwoeq} in case of green electricity are in contrast with \SI{16606}{\tcotwoeq} for the University of Freiburg. The main difference results from the emissions for heating/cooling, where district heating for the Leibniz University Hannover generates significantly fewer emissions than the heating/cooling used at the University of Freiburg. In the latter case, the usage of ground water for cooling is specified, but no information on the type of heating by the University of Freiburg is provided.

The number of students at the Leibniz University Hannover is given in the environmental report 2017--2019~\cite{UniHann} as \SI{29781}{}, and the number of employees as \SI{4948}{}, summing to \SI{34729}{} members of the Leibniz University Hannover. This results in emissions per person per year of \SI{0.23}{\tcotwoeq} (\SI{0.87}{\tcotwoeq}) with green (conventional) electricity supply, with the caveat of not accounting for procurement.

Given the lack of information on procurement for the Leibniz University Hannover, the larger emissions for heating/cooling by the University of Freiburg and approximate similarity in the required other numbers, the values obtained for the University of Freiburg are used as conservative estimate for the institutional footprint of a university.

\section{Emissions from procurement}
\label{app:procurement}

Procurement is a major source of emissions, in particular for organisations such as a university, whose core functionalities: research, teaching and knowledge transfer, can be assigned to the services sector.

\begin{figure}
    \centering
    \includegraphics[width=0.9\textwidth]{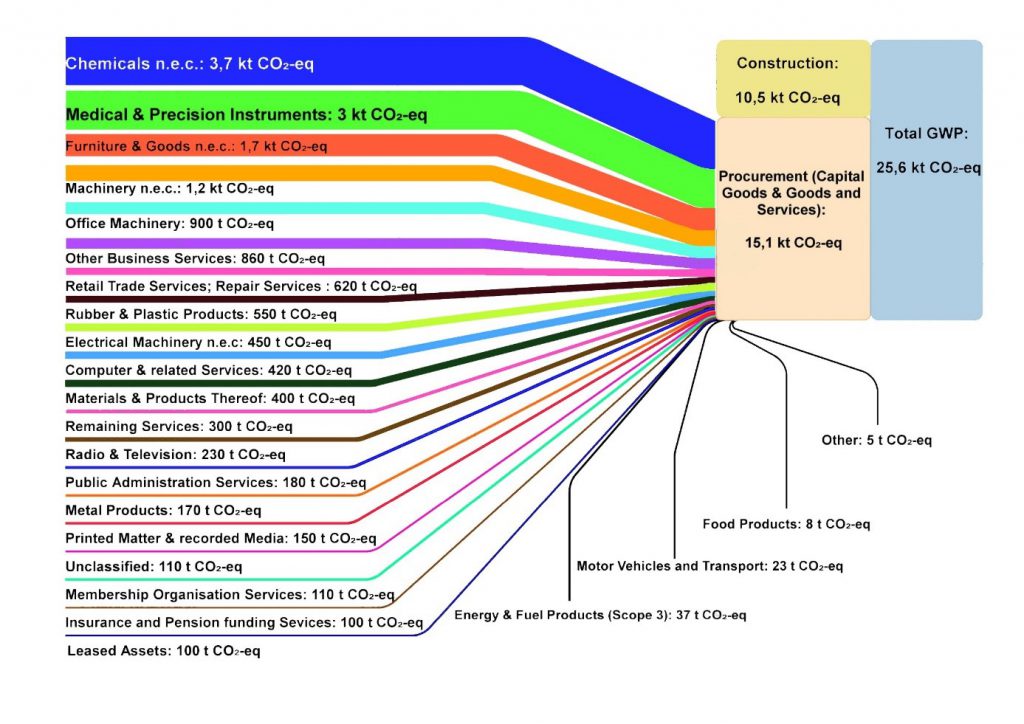}
    \caption{Distribution of scope 3 greenhouse gas emissions of the two categories \emph{Goods and services} and \emph{Capital goods} for the University of Freiburg in 2017~\cite{ProcurementUniFR}. \emph{Construction} is listed as separate scope 3 category for comparison, but it is unclear which category in the environmental report 2019/2020~\cite{UniFR} this has been added to.}
    \label{fig:ProcurementUniFR}
\end{figure}

The procurement-related emissions by the University of Freiburg, discussed in the paper and obtained from the environmental report 2019/2020~\cite{UniFR}, are based on procurement data from 2017, analysed in a master thesis~\cite{ProcurementMasterThesis}, which is unfortunately not available online. A subsequent report is accessible in Ref.~\cite{ProcurementUniFR}, where the relevant figure has been regenerated for the report. Figure~\ref{fig:ProcurementUniFR} displays the scope 3 procurement distribution of the categories \emph{Goods and services} and \emph{Capital goods}, contained in Ref.~\cite{ProcurementUniFR}. The combined emissions listed as \SI{15100}{\tcotwoeq} do not exactly equal the value of \SI{14486}{\tcotwoeq}, quoted in the environmental report 2019/2020 by the University of Freiburg~\cite{UniFR}, but are very close. The displayed split-up can therefore be seen as representative of the value, quoted in the environmental report 2019/2020.

Largest contributors to the emissions from procurement are the categories: \emph{Chemicals} (\SI{3700}{\tcotwoeq}), \emph{Medical \& Precision Instruments} (\SI{3000}{\tcotwoeq}), \emph{Furniture \& Goods n.e.c.} (\SI{1700}{\tcotwoeq}), and \emph{Machinery n.e.c.} (\SI{1200}{\tcotwoeq}), which amount to about \SI{64}{\%} of the procurement-related emissions. 
The large number of categories gives an idea of how difficult it is to address the reduction of procurement-related emissions; due to its large overall contribution, it remains critical though. Two approaches are possible for this purpose: demand management by the university, and \emph{green} procurement. The choice of the best approach might differ depending on the category.

\begin{figure*}
    \centering
    \includegraphics[width=0.9\textwidth]{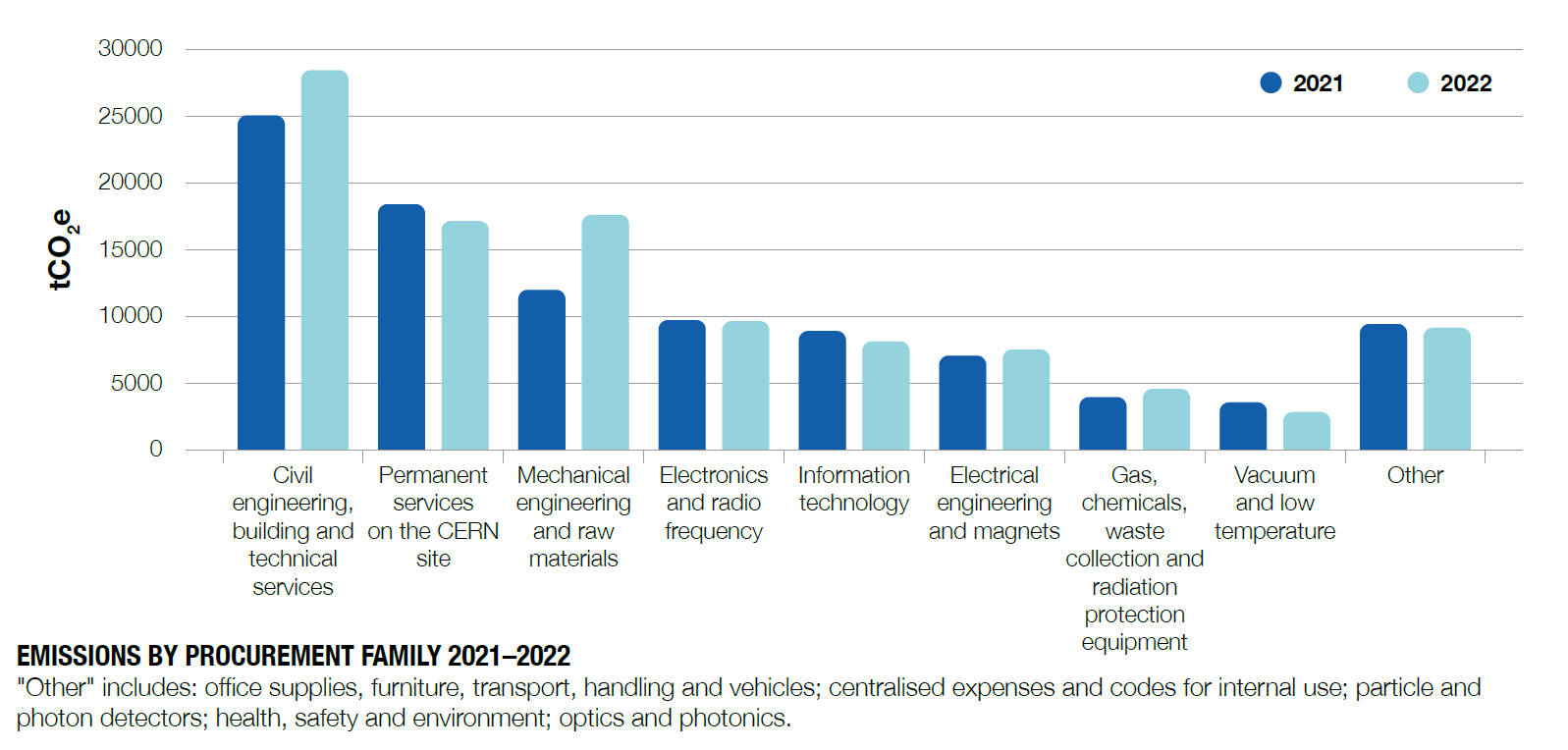}
    \caption{Emissions by procurement family 2021--2022~\cite{CERN2122}. The category \emph{Other} includes office supplies, furniture, transport, handling and vehicles; centralised expenses and codes for internal use; particle and photon detectors; health, safety and environment; optics and photonics.}
    \label{fig:ProcurementCERN}
\end{figure*}

Details on procurement-related emissions by CERN are provided in the environmental report 2021--2022~\cite{CERN2122}. The figure \emph{Emissions by procurement family 2021--2022} from page 23 of the CERN environmental report 2021--2022~\cite{CERN2122} is reproduced in Figure~\ref{fig:ProcurementCERN}.

The largest contributors to procurement-related emissions in 2022 are the following categories: \emph{Civil engineering, building and technical services}, \emph{Permanent services on the CERN site}, and \emph{Mechanical engineering and raw materials}. These could be related to construction regarding buildings or machinery, which can be seen either as part of the infrastructure of the accelerator complex, usual construction and maintenance of buildings of an organisation, or engineering work, which could also be performed at a university. It is thus not possible to explicitly assign the procurement categories to the footprints of accelerators, experiments, upgrade activities, or CERN as an institute. 

A direct comparison of the procurement categories to those of the University of Freiburg is not possible.

\section{Emissions for various transportation modes}
\label{app:TransportModes}

\begin{figure}[h]
    \centering
    \includegraphics[width=0.75\textwidth,trim={5cm 0 5cm 0},clip]{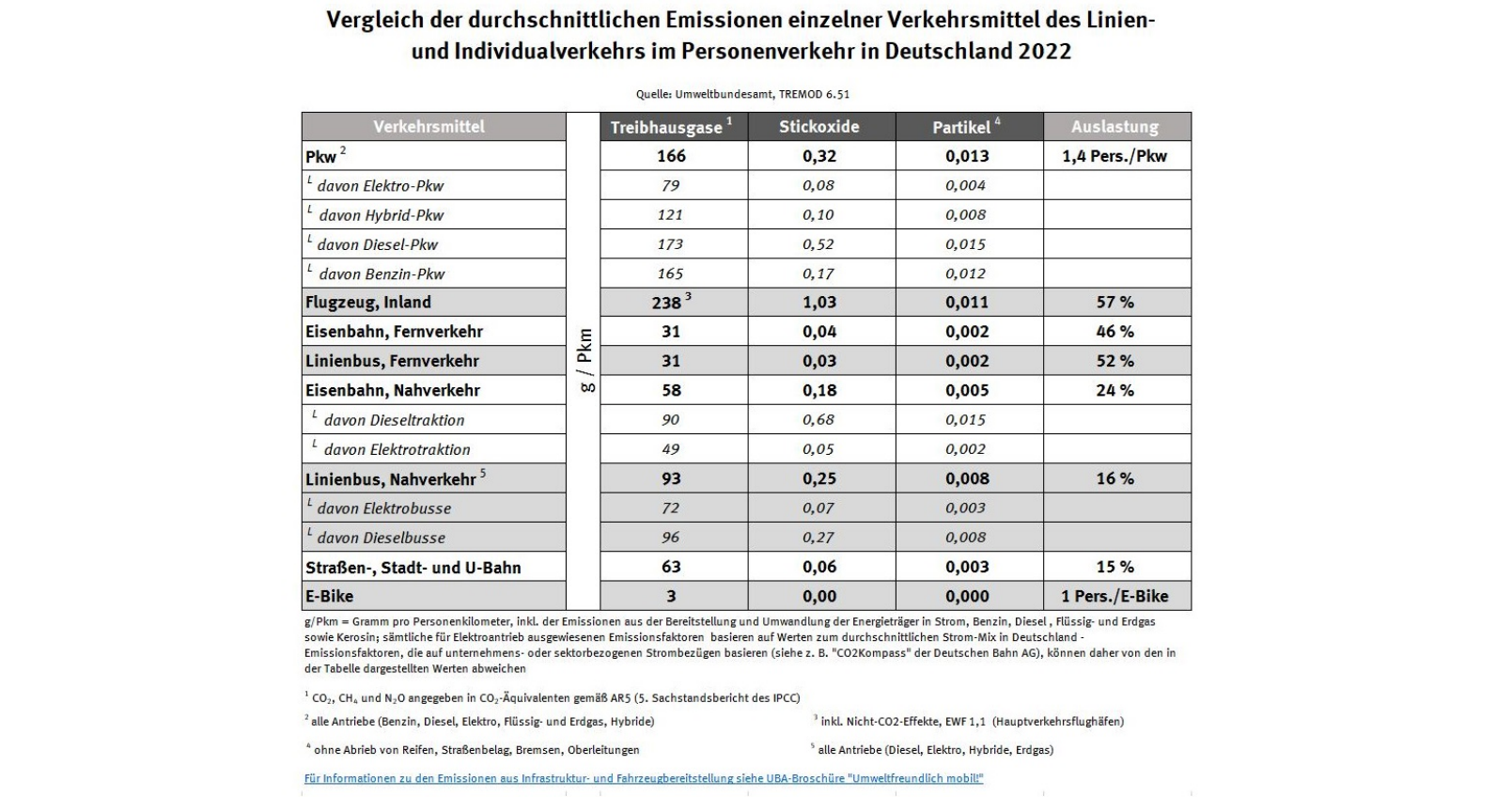}
    \caption{Comparison of average \cotwo-equivalent emissions for various modes of transport in Germany for 2022 listed in \SI{}{g\cotwo e}~\cite{Umweltbundesamt_TREMOD}. For the first version of the Kyf calculator, only long-distance trains (\emph{Eisenbahn, Fernverkehr}), long-distance buses (\emph{Linienbus, Fernverkehr}) and personal car (\emph{Pkw}) were used.}
    \label{fig:Transport}
\end{figure}

The German Federal Environment Agency (\emph{Umweltbundesamt} - UBA) provides a comparison of \cotwo emissions for various modes of transport per person per kilometre in a table on their webpage~\cite{Umweltbundesamt_TREMOD}. This table represents the data collected for 2022 and is reproduced in Figure~\ref{fig:Transport}. For the calculation of the average emission values, an average occupancy is also estimated to obtain a comparable value for an individual across all modes of transport.

For the first version of the Kyf calculator, only long-distance trains (\emph{Eisenbahn, Fernverkehr}), long-distance buses (\emph{Linienbus, Fernverkehr}) and personal car (\emph{Pkw}) were considered. The \cotwo emissions for intra-Europe and transcontinental flights were obtained from Ref.~\cite{UBAcalculator} in order to have the emission values corresponding to the duration of the flight rather than the distance covered by it. This was considered to be more user-friendly for the Kyf calculator. Other modes of transportation listed in Figure~\ref{fig:Transport} are primarily relevant for short-distance commute and were therefore not incorporated in the first version of the Kyf calculator. These could potentially be included in future iterations.

\end{document}